\documentclass[a4paper,UKenglish,cleveref, autoref]{lipics-v2019}

\title{One-dimensional guarded fragments}

\author{Emanuel Kiero\'nski}{University of Wroc\l{}aw, Poland}{kiero@cs.uni.wroc.pl}{https://orcid.org/0000-0002-8538-8221}{}

\authorrunning{E. Kiero\'nski}

\Copyright{E. Kiero\'nski}



\category{}

\relatedversion{}

\supplement{}

\funding{Supported by {Polish National Science Centre}{} grant No {2016/21/B/ST6/01444}{}.}

\acknowledgements{}

\EventEditors{Peter Rossmanith, Pinar Heggernes, and Joost-Pieter Katoen}
\EventNoEds{3}
\EventLongTitle{44th International Symposium on Mathematical Foundations of Computer Science (MFCS 2019)}
\EventShortTitle{MFCS 2019}
\EventAcronym{MFCS}
\EventYear{2019}
\EventDate{August 26--30, 2019}
\EventLocation{Aachen, Germany}
\EventLogo{}
\SeriesVolume{138}
\ArticleNo{5}
\nolinenumbers 
\hideLIPIcs  

\begin{document}

\maketitle


\newcommand{\cT}{\mathcal{T}}
\newcommand{\cL}{\mathcal{L}}
\newcommand{\cF}{\mathcal{F}}

\newcommand{\phie}{\phi^{ \exists}}
\newcommand{\phiu}{\phi^{\sss \forall}}

\newcommand{\psie}{\psi^{\sss \exists}}
\newcommand{\psiu}{\psi^{\sss \forall}}
\newcommand{\mse}{m_{\sss \exists}}
\newcommand{\msu}{m_{\sss \forall}}

\newcommand{\tpstar}{{\rm tp}^*}
\newcommand{\tp}{{\rm tp}}
\newcommand{\UF}{\mbox{$\mbox{\rm UF}_1$}}
\newcommand{\UFAm}{\mbox{$\mbox{\rm UF}_1^{\sss Alt^-}$}}
\newcommand{\UFA}{\mbox{$\mbox{\rm UF}_1^{\sss Alt}$}}

\newcommand{\ODF}{\mbox{$\mbox{\rm F}_1$}}

\newcommand{\SUF}{\mbox{$\mbox{\rm sUF}_1^{\sss =}$}}

\newcommand{\UFEQ}{\mbox{$\mbox{\rm UF}_1^{\sss =}(\sim)$}}
\newcommand{\SUFEQ}{\mbox{$\mbox{\rm sUF}_1^{\sss =}(\sim)$}}

\newcommand{\ST}{L}
\newcommand{\ZR}{R_0}
\newcommand{\NR}{R_2}
\newcommand{\FOtDTC}{\mbox{$\mbox{\rm FO}^2+\mbox{$\mbox{\rm DTC}$}(E)$}}
\newcommand{\FOtDTCp}{\mbox{$\mbox{\rm FO}^2+\mbox{$\mbox{\rm DTC}$}(E)+[*]$}}

\newcommand{\DT}{\overline{E}}


\newcommand{\UFO}[3]{
\ifthenelse{\equal{#3}{111}}
{$\forall^#1_{TC}[#2]$}{$\forall^#1_{TC}[#2,#3]$}}

\newcommand{\DDD}{\mbox{\large \boldmath $\delta$}}

\newcommand{\UtTCo}{\mbox{$\mbox{\rm UTC}^2_1$}}
\newcommand{\UtTCt}{\mbox{$\mbox{\rm UTC}^2_2$}}
\newcommand{\UthTCo}{\mbox{$\mbox{\rm UTC}^3_1$}}

\newcommand{\bp}{\mathbf{p}}
\newcommand{\br}{\mathbf{r}}
\newcommand{\bP}{\mathbf{P}}
\newcommand{\bR}{\mathbf{R}}
\newcommand{\bE}{\mathbf{E}}
\newcommand{\bL}{\mathbf{L}}
\newcommand{\bO}{\mathbf{O}}
\newcommand{\bU}{\mathbf{U}}
\newcommand{\bgeq}{\mathbf{\succeq}}%

\newcommand{\cA}{\mathcal{A}}
\newcommand{\cB}{\mathcal{B}}
\newcommand{\cC}{\mathcal{C}}
\newcommand{\cD}{\mathcal{D}}
\newcommand{\cE}{\mathcal{E}}
\newcommand{\cP}{\mathcal{P}}
\newcommand{\cQ}{\mathcal{Q}}

\newcommand{\ord}{\mbox{\rm ord}}
\renewcommand{\deg}{\mbox{\rm deg}}
\newcommand{\Deg}{\mbox{\rm Deg}}
\newcommand{\DEG}{\mbox{\rm DEG}}
\newcommand{\NPTime}{\mbox{\sc{NPTime}}}%

\newcommand{\lrLink}{+}%
\newcommand{\rlLink}{-}%
\newcommand{\bdLink}{\circ}%


\newcommand{\bC}{\mathbf{C}}%

\newcommand{\cK}{\mathcal{K}}%
\newcommand{\cG}{\mathcal{G}}
\newcommand{\bbP}{\mathbb{P}}
\newcommand{\fA}{\mathfrak{A}}%
\newcommand{\fB}{\mathfrak{B}}%
\newcommand{\fC}{\mathfrak{C}}%

\newcommand{\fD}{\mathfrak{D}}%

\renewcommand{\phi}{\varphi} 

\newcommand{\zz}{\mathit{\!0\!0}}
\newcommand{\zo}{\mathit{\!0\!1}}
\newcommand{\oz}{\mathit{1\!0}}
\newcommand{\oo}{\mathit{1\!1}}

\newcommand{\AAA}{\mbox{\large \boldmath $\alpha$}}
\newcommand{\BBB}{\mbox{\large \boldmath $\beta$}}

\newcommand{\EQ}{\ensuremath{{\mathcal EQ}}}
\newcommand{\Sat}{\ensuremath{\textit{Sat}}}
\newcommand{\FinSat}{\ensuremath{\textit{FinSat}}}

\newcommand{\PDL}{\mbox{\rm PDL}}
\newcommand{\FO}{\mbox{\rm FO}}
\newcommand{\FOt}{\mbox{$\mbox{\rm FO}^2$}}
\newcommand{\FOthree}{\mbox{$\mbox{\rm FO}^3$}}

\newcommand{\FOtEC}{\mbox{$\mbox{\rm EC}^2$}}
\newcommand{\FOtECth}{\mbox{$\mbox{\rm EC}^2_3$}}
\newcommand{\FOtECt}{\mbox{$\mbox{\rm EC}^2_2$}}
\newcommand{\FOtECo}{\mbox{$\mbox{\rm EC}^2_1$}}
\newcommand{\FOtECk}{\mbox{$\mbox{\rm EC}^2_k$}}
\newcommand{\GFOTC}{\mbox{$\mbox{\rm G}_\exists \mbox{\rm FO}^2+\mbox{\rm TC}$}}

\newcommand{\GgtwoF}{\mbox{$\mbox{\rm TGF}$}}

\newcommand{\GFonedim}{\mbox{$\mbox{\rm GF}_1$}}
\newcommand{\LGFonedim}{\mbox{$\mbox{\rm LGF}_1$}}

\newcommand{\GgtwoFonedim}{\mbox{$\mbox{\rm TGF}_1$}}

\newcommand{\UNFO}{\mbox{\rm UNFO}}
\newcommand{\LGF}{\mbox{\rm LGF}}

\newcommand{\GNFO}{\mbox{\rm GNFO}}

\newcommand{\GF}{\mbox{\rm GF}}

\newcommand{\BS}{\mbox{\rm BS}}

\newcommand{\GFt}{\mbox{$\mbox{\rm GF}^2$}}

\newcommand{\FOtEth}{\mbox{$\mbox{\rm EQ}^2_3$}}
\newcommand{\FOtEt}{\mbox{$\mbox{\rm EQ}^2_2$}}
\newcommand{\FOtEo}{\mbox{$\mbox{\rm EQ}^2_1$}}
\newcommand{\FOtEk}{\mbox{$\mbox{\rm EQ}^2_k$}}

\newcommand{\NLogSpace}{\textsc{NLogSpace}}
\newcommand{\LogSpace}{\textsc{LogSpace}}
\newcommand{\NP}{\textsc{NPTime}}
\newcommand{\PTime}{\textsc{PTime}}
\newcommand{\PSpace}{\textsc{PSpace}}
\newcommand{\ExpTime}{\textsc{ExpTime}}
\newcommand{\ExpSpace}{\textsc{ExpSpace}}
\newcommand{\NExpTime}{\textsc{NExpTime}}
\newcommand{\TwoExpTime}{2\textsc{-ExpTime}}
\newcommand{\TwoNExpTime}{2\textsc{-NExpTime}}
\newcommand{\APSpace}{\textsc{APSpace}}
\newcommand{\AExpSpace}{\textsc{AExpSpace}}
\newcommand{\ASpace}{\textsc{ASpace}}
\newcommand{\DTime}{\textsc{DTime}}

\newcommand{\set}[1]{\{#1\}}
\newcommand{\md}[2][] {{\lfloor#2\rfloor_{#1}}}
\newcommand{\sizeOf}[1]{\lVert #1 \rVert}
\newcommand{\str}[1]{{\mathfrak{#1}}}
\newcommand{\restr}{\!\!\restriction\!\!}

\newcommand{\N}{{\mathbb N}}
\newcommand{\Z}{{\mathbb Z}}





\newcommand{\Racl}{r_1^{\scriptscriptstyle \#}}
\newcommand{\Rbcl}{r_2^{\scriptscriptstyle \#}}
\newcommand{\Ricl}{r_i^{\scriptscriptstyle \#}}
\newcommand{\Rkcl}{r_k^{\scriptscriptstyle \#}}
\newcommand{\Raclp}{{r_1'}^{\scriptscriptstyle \#}}
\newcommand{\Rbclp}{{r_2'}^{\scriptscriptstyle \#}}
\newcommand{\Raclb}{s_1}
\newcommand{\Rbclb}{s_2}
\newcommand{\Riclb}{s_i}

\newcommand{\taucl}{\tau^{\scriptscriptstyle \#}}

\newcommand{\Eo}{E_1^{}}
\newcommand{\Et}{E_2^{}}
\newcommand{\Ei}{E_i^{}}

\newcommand{\sss}{\scriptscriptstyle}
\newcommand{\emodels}{\models_{\sss \#}}
\newcommand{\pmodels}{\mid \hspace*{-1.5pt}\approx_{\sss \#}}
 




\newtheorem{openquestion}{\bf Open Question}

\newcommand{\K}{K}


\newcommand{\ODW}{\mbox{$\mbox{\rm F}_1[<, +1]$}}

\newcommand{\sep}{\iota}
\newcommand{\succs}{'+'}
\newcommand{\lesss}{'-'}
\newcommand{\prof}[3]{{\rm prof}^{\str{#1}}_{#2}({#3})}

\newcommand{\Qfr}{\mbox{Q}}

\newcommand{\type}[2]{{\rm tp}^{{#1}}({#2})}

\newcommand{\lambdanext}{\lambda^{\sss +1}}
\newcommand{\lambdadiff}{\lambda^{\sss \not=}}
\newcommand{\lambdaless}{\lambda^{\sss <}}

\begin{abstract}

We call a first-order formula one-dimensional if every maximal block of existential (or universal) quantifiers in it leaves  at most one variable free.
We consider the one-dimensional restrictions of the guarded fragment, \GF, and the tri-guarded fragment, \GgtwoF{},
the latter being a recent extension of \GF{} in which quantification for subformulas with at most two free variables need not be guarded, and which thus may 
be seen as a unification of \GF{} and the  two-variable fragment, \FOt. We denote the
resulting formalisms, resp., \GFonedim, and \GgtwoFonedim{}. We show that \GFonedim{} has an exponential model property and
\NExpTime-complete satisfiability problem (that is, it is easier than full \GF{}). For \GgtwoFonedim{} we show that it is decidable, has the finite model property, and its satisfiability problem is  \TwoExpTime-complete (\NExpTime-complete in the absence of equality).
All the above-mentioned results are obtained for signatures with no constants. We finally discuss the impact of their addition, observing  that constants do not spoil the decidability but 
increase the complexity of the satisfiability problem.
\end{abstract}

\section{Introduction}

The \emph{guarded fragment} of first-order logic, \GF{}, is obtained by requiring all quantifiers to be appropriately relativised by atoms. It was introduced
by Andr{\'e}ka,  van Benthem and N\'{e}meti \cite{ABN98} as a 
generalization of propositional modal logic and may be also seen as an extension of some standard description logics.
\GF{} has good algorithmic and model-theoretic properties. In particular, 
Gr\"adel proved that its  satisfiability problem is decidable, it has a tree-like model property and the finite model property \cite{Gra99}. 
The idea of \GF{} turned out to be very fruitful and found numerous applications.  
In this paper we consider some modifications of the syntax of \GF{}. Our aim is to check if in this way we can obtain interesting fragments with better complexity 
and/or  attractive expressiveness.

The satisfiability problem for \GF{} is \TwoExpTime-complete. This relatively high complexity can be lowered to \ExpTime{} either by
bounding the number of variables, or the arity of relation symbols \cite{Gra99}. We propose another way of decreasing the complexity 
without sacrificing either the number of variables or the arity of relations. The idea is to restrict formulas to be \emph{one-dimensional}. 
We say that a formula
is one-dimensional if  every maximal block of existential (or universal) quantifiers in it leaves at most one variable free.
We remark that the one-dimensional restriction of full first-order logic, \ODF, is undecidable, as observed by Hella and Kuusisto \cite{HK14}. We denote the
intersection of \ODF{} and \GF{}  by \GFonedim{}  and call it the \emph{one-dimensional guarded fragment}. While this variation decreases the expressive power of the logic, we believe that it is still quite interesting, as, in particular, it still embeds propositional modal logic, and most standard
description logics embeddable in full \GF{}. Thus, as \GF{}, it may serve as an extension of modal/description logics to contexts with relations of arbitrary arity.
We show that the satisfiability problem for \GFonedim{} is \NExpTime-complete and that it has an exponential model property, that is, 
its every satisfiable formula has a model of size bounded exponentially in its length. This is in contrast to full \GF{} in which one can enforce
doubly exponentially large models.
Moreover, proving the finite model property for \GFonedim{} is much easier than for full \GF{},
in particular it does not need complicated combinatorial constructions used in the case of \GF{} (in \cite{Gra99}, and in 
B{\'{a}}r{\'{a}}ny, Gottlob and Otto \cite{BGO14}).
We obtain a corresponding  \NExpTime-lower bound  even for a weaker logic,  \emph{uniform} \GFonedim, that is the intersection of 
\GFonedim{} and \emph{uniform} \ODF{}, \UF,
the latter being a decidable restriction of \ODF{} introduced in \cite{HK14} as a canonical generalization of the two-variable fragment \FOt{} (with equality) to 
scenarios involving relations
of arity greater than two (see Kiero\'nski, Kuusisto \cite{KK14} where \NExpTime-completeness of \UF{} is shown). This is slightly surprising, since in many aspects \UF{} behaves similarly to the two-variable fragment, \FOt{}, and the guarded version of the latter is \ExpTime-complete \cite{Gra99}.

We also consider an extension of \GF{} called the \emph{tri-guarded fragment}, \GgtwoF. 
In \GgtwoF{} quantification for subformulas with at most two free variables may  be used freely, 
without guards.  Hence, \GgtwoF{}  unifies \GF{} and the already-mentioned  \FOt{}. 
We borrowed the term \emph{tri-guarded fragment} from a recent work by
Rudolph and \v{S}imkus \cite{RS18}, but, actually, the idea behind TGF is not 
new and can be traced back already in Kazakov’s PhD thesis \cite{Kaz06}
where the fragment \GF|\FOt{}, essentially identical with TGF, was defined.
A similar logic, \GF{} \emph{with binary cross product}, \GF$^{\times_2}$,
is also considered by Bourhis, Morak and Pieris \cite{BMP17}. Both \GF|\FOt{} and
\GF$^{\times_2}$  do not allow constant 
symbols. We remark that in our initial scenario we also assume that  constants are not present in signature; however, we will discuss their addition later.

Similarly to \GF{}, \FOt{} is a seminal fragment of first-order logic, and its importance is justified, \emph{inter alia}, by its close relationships to modal and description logics. Mortimer \cite{Mor75} demonstrated that it has the finite model property  and Gr\"adel, Kolaitis and Vardi  \cite{GKV97} proved  that its  satisfiability problem is \NExpTime-complete.
Each of the logics \GF, \FOt{} has some advantages and drawbacks with respect to the other. We mention here the fact that \GF{} allows only to express properties of a local character, \emph{e.g.}, it cannot express $\forall xy (Px \wedge Qy \rightarrow Rxy)$, while \FOt{} does not allow for a non-trivial use of relations of arity greater than two. 
\GgtwoF{} offers a substantial improvement in these aspects. Moreover, in  \GgtwoF{} we can embed the G\"odel class, that is 
the class of all prenex formulas of the form $\forall xy \exists \bar{z} \psi(x,y,\bar{z})$. Indeed, any such formula has an equisatisfiable \GgtwoF{} formula
obtained  just by an addition of a dummy guard, as follows, $\forall xy \exists \bar{z} (G(x,y,\bar{z}) \wedge \psi(x,y,\bar{z}))$, where $G$ is a fresh relation symbol of the appropriate arity.
Such embedding implies, however,  that the satisfiability problem for \GgtwoF{} with equality is undecidable, since the G\"odel class with equality 
is undecidable, as proved by Goldfarb \cite{Gol84}. 
The undecidability of \GgtwoF{} with equality is also shown  
in
\cite{RS18} by a direct grid encoding.
On the positive side, it turns out that the satisfiability problem for  \GgtwoF{} \emph{without equality} is decidable and \TwoExpTime-complete.
It was proved in \cite{Kaz06} by a resolution method, and follows also from the decidability of \GF$^\times_2$, 
shown in \cite{BMP17} by  a use of the database-theoretic concept of  \emph{chase}.\footnote{A footnote 
in \cite{BMP17} suggests that the decidability of \GF{} with binary cross-product is retained in the presence of equality. This has however been later later refuted by the authors (private
communication). \GF{} with binary cross product with equality is undecidable by the same arguments we gave for \GgtwoF.}

In this paper we  consider a natural combination of \GFonedim{} and \GgtwoF{}, the \emph{one-dimensional tri-guarded fragment}, \GgtwoFonedim{}, which,
on the one hand,  allows us to use unguarded quantification for subformulas with at most two free variables, but, on the other hand,
requires to obey the one-dimensionality restriction. We show that this variant is decidable even in the presence of equality. The complexity, however, depends on the presence/absence of
equality: The satisfiability problem is \TwoExpTime-complete with equality and \NExpTime-complete without it. The logic has the finite model property (we remark that
whether full \GgtwoF{} has the finite model property is an open question), and, again, a bound on the  size of minimal models is
doubly- or singly exponential, depending on whether equality is allowed or not.  
\GgtwoFonedim{} may be seen 
as a decidable generalization of \FOt{} (with equality) to scenarios with relations of arity greater than two, 
alternative and orthogonal in the expressive power to the above-mentioned \UF{}. We also remark that \GgtwoFonedim{} can express the concept
of \emph{nominals} from description logics, since the combination of equality and unguarded quantification for subformulas with two free variables allows us to say that some
unary predicates hold for unique elements of a model. Thus we can embed in \GgtwoFonedim{}, \emph{e.g.}, the description logic $\mathcal{ALC}$ plus inverse roles ($\mathcal{I}$), nominals ($\mathcal{O}$),
role hierarchies ($\mathcal{H}$), and any Boolean combination of roles (including their negations).

We then briefly consider applications of the ideas of one-dimensionality and tri-guardedness to two decidable
extensions of \GF, namely, the loosely guarded fragment, \LGF{}, introduced by van Benthem  \cite{Ben97}, and the guarded negation fragment, \GNFO{}, proposed by B{\'{a}}r{\'{a}}ny, ten Cate and Segoufin \cite{BtCS15}. 
Regarding one-dimensionality, it helps in the case of \LGF: one-dimensional \LGF{} has an exponential model property
and \NExpTime-complete satisfiability problem (exactly as \GFonedim), but does not help in the case of \GNFO{}, where the one-dimensional 
variant remains \TwoExpTime-hard. Regarding the tri-guardedness, the results are negative:
both \LGF{} and \GNFO{}, even in their one-dimensional variants, become undecidable when unguarded quantification for subformulas with two
free variables is allowed.

As remarked, all the results discussed above are obtained under the assumption that constants are not present in signatures. 
It turns out that all the decidability results are preserved in the presence of constants. However, interestingly,
the computational complexity may change (we recall that for \GF{} constants make no difference \cite{Gra99}). 
This is also the case for \GgtwoF{}  with constants, without equality, which  is shown in \cite{RS18} to be \TwoNExpTime-complete. Here we show that a \TwoNExpTime-lower bound can be obtained 
even for \GgtwoFonedim{}   with constants, without equality. We also observe that the presence of constants lifts the complexity of  
 \GFonedim{} to \TwoExpTime{}.

In Table \ref{tab:complexitiesconst} we summarize the above-discussed complexity results for the variations of \GF.
We point out an interesting status of \GgtwoFonedim{}: it is 
\NExpTime-complete without equality and constants, \TwoExpTime-complete with equality and without constants, and \TwoNExpTime-complete
with constants (with or without equality).

We finally remark that 
 further pushing the concepts of one-dimensionality and tri-guardedness to, resp., \emph{two}-dimensionality and 
\emph{tetra}-guardedness does not lead to attractive results. Indeed, a \TwoExpTime{} lower bound for two-dimensional \GF{} can be shown by
a slight adaptation of the bound for full \GF{} from \cite{Gra99};
 allowing for unguarded quantification for subformulas with three free variables gives undecidability, as the
resulting logic contains the undecidable three-variable fragment  of \FO{} (see, \emph{e.g.},  Kahr, Moore and Wang \cite{KMW62}).
Undecidability of the three-variable fragment can be easily shown even using only one-dimensional formulas. 

\begin{table}
\begin{center}
\begin{tabular}{|c|c|c|c|c|} \hline
logic & with $=$ & without =  \\ \hline  
\GF{} & \TwoExpTime. & \TwoExpTime\\  
\GgtwoF{} & undecidable & \TwoExpTime{} (\TwoNExpTime) \\  
\GFonedim{}& {\bf \NExpTime{} (\TwoExpTime{})} & {\bf \NExpTime{} (\TwoExpTime{})} \\ 
\GgtwoFonedim{} & {\bf \TwoExpTime{} (\TwoNExpTime{})} & {\bf \NExpTime{} (\TwoNExpTime{})}\\ 
\hline
\end{tabular}
\end{center}
\caption{Complexities of the guarded fragments. If the presence of constants makes a difference, the complexity of the variant with constants is given in the brackets. All logics have the finite model property. Results of this paper are distinguished in bold.}\label{tab:complexitiesconst}
\end{table}

\section{Preliminaries} \label{s:prel}

We mostly work with purely relational signatures with no constants and  function symbols (only in Section \ref{s:constants} we consider signatures with constants). For convenience we also assume that there are no relation symbols of arity $0$.
We refer to structures using Fraktur capital letters, and to their domains using
the corresponding Roman capitals. Given a structure $\str{A}$ and some $B \subseteq A$ we
denote by $\str{A} \restr B$ or just by $\str{B}$ the restriction of $\str{A}$ to its subdomain $B$. 

We usually use $a, b, \ldots$ to denote elements from domains of structures, $\bar{a}$, $\bar{b}$, $\ldots$ for tuples of elements, $x$, $y$, $\ldots$ for
variables and $\bar{x}$, $\bar{y}$, $\ldots$ for tuples of variables; all of these possibly with some decorations.
For a tuple of variables $\bar{x}$ we use $\psi(\bar{x})$ to denote  a formula (or subformula) $\psi$,
whose all free variables are in $\bar{x}$.

An {\em atomic} $l$-{\em type} $\beta$ over a signature $\sigma$ is a
maximal consistent set of atomic or negated atomic formulas (including equalities/inequalities) over
$\sigma$ in $l$ variables $x_1, \ldots, x_l$. We often identify a type with the conjunction of its elements, $\beta(x_1, \ldots, x_l)$. 
For an $l$-type $\beta$ we denote by $\beta \restr x_i$ ($i=1, \ldots, l$) the $1$-type obtained
by removing from $\beta$ all the literals that use some $x_j$, with $j \not=i$, and then replacing all occurrences of $x_i$ by $x_1$.
We will be particularly interested in $1$-types and $2$-types over signatures $\sigma$ consisting of the
relation symbols used in some given formula. Observe that the number of $1$-types is bounded by a function which is exponential in $|\sigma|$, and hence also in the length
of the formula. This is because a $1$-type just corresponds to a subset of $\sigma$. On the other hand, the number of $2$-types 
may be doubly exponentially large. Indeed, using an $n$-ary predicate and two fixed variables one can build $2^n$ atoms which
then can be used to form $2^{2^n}$ different $2$-types.

Let $\str{A}$ be a structure, and let
$a,b\in A$ be such that $a\neq b$. We denote by $\type{\str{A}}{a}$ the unique atomic
1-type \emph{realized} in $\str{A}$ by the element $a$, \emph{i.e.}, the $1$-type $\alpha(x)$ such that $\str{A} \models \alpha(a)$; similarly by
$\type{\str{A}}{a,b}$ we denote the unique atomic 2-type realized in $\str{A}$ by
pair $(a,b)$, \emph{i.e.}, the $2$-type $\beta(x,y)$ such that $\str{A} \models \beta(a,b)$.   
For $B \subseteq A$ we denote by $\AAA[B]$  the set of all $1$-types realized in $\str{A}$ by elements of $B$. 

Below we define several fragments of first-order logic, \FO{}, including two new fragments, \GFonedim{} and \GgtwoFonedim{}. Each of the fragments is defined as the least set
of formulas (i) containing all atomic formulas (including equalities), (ii) closed under 
Boolean connectives, and (iii) satisfying appropriate (depending on the fragment) rules of using quantifiers, specified below
($\bar{x}$, $\bar{y}$ represent here any tuples of variables and $x$, $y$ represent any variables):

\begin{itemize}
\item \emph{Guarded fragment} of first-order logic, \GF: 
\begin{itemize}
\item
if $\psi(\bar{x}, \bar{y}) \in$ \GF{} then $\forall \bar{x}  (\gamma(\bar{x}, \bar{y}) \rightarrow 
\psi(\bar{x},\bar{y}))$ and $\exists \bar{x}  (\gamma(\bar{x},\bar{y}) \wedge \psi(\bar{x},\bar{y}))$
belong to \GF{}, where $\gamma(\bar{x},\bar{y})$ is an atomic formula containing 
all the free variables of $\psi$, called a \emph{guard} for $\psi$. 
\end{itemize}

\item \emph{One-dimensional fragment} of first-order logic, \ODF:
\begin{itemize}
\item
if $\psi(\bar{x},y) \in$ \ODF{} then $\exists \bar{x} \psi(\bar{x},y)$ and
$\forall \bar{x} \psi(\bar{x},y)$ belong to \ODF{}.
\end{itemize}

\item \emph{One-dimensional guarded fragment}, \GFonedim{}:
\begin{itemize}
\item
if $\psi(\bar{x}, y) \in$ \GFonedim{} then $\forall \bar{x}  (\gamma(\bar{x}, {y}) \rightarrow 
\psi(\bar{x},{y}))$ and $\exists \bar{x}  (\gamma(\bar{x},{y}) \wedge \psi(\bar{x},{y}))$ belong to \GFonedim{},
where $\gamma(\bar{x},{y})$ is a guard for $\psi$.
\end{itemize}

\item \emph{Tri-guarded  fragment}, \GgtwoF:
\begin{itemize}
\item 
 if $\psi(\bar{x}, \bar{y}) \in$ \GgtwoF{} then $\forall \bar{x}  (\gamma(\bar{x}, \bar{y}) \rightarrow 
\psi(\bar{x},\bar{y}))$ and $\exists \bar{x}  (\gamma(\bar{x},\bar{y}) \wedge \psi(\bar{x},\bar{y}))$
belong to \GgtwoF{}, where $\gamma(\bar{x},\bar{y})$ is a guard for $\psi$,
\item if
$\psi(x,y)$ is in \GgtwoF, then $\exists x \psi(x,y)$ and $\forall x \psi(x,y)$ belong to \GgtwoF{}.
\end{itemize}

\item  \emph{One-dimensional tri-guarded fragment}, \GgtwoFonedim:
\begin{itemize}
\item
if $\psi(\bar{x}, y) \in$ \GgtwoFonedim{} then $\forall \bar{x}  (\gamma(\bar{x}, {y}) \rightarrow 
\psi(\bar{x},{y}))$ and $\exists \bar{x}  (\gamma(\bar{x},{y}) \wedge \psi(\bar{x},{y}))$ belong to \GgtwoFonedim{},
where $\gamma(\bar{x},{y})$ is a guard for $\psi$,
\item if
$\psi(x,y)$ is in \GgtwoFonedim, then $\exists x \psi(x,y)$ and $\forall x \psi(x,y)$ belong to \GgtwoFonedim{}.

\end{itemize}

\end{itemize}

Note that \GFonedim{} is just the intersection of \GF{} and \ODF{},
\GgtwoF{} contains both \GF{} and  \FOt{}, and
\GgtwoFonedim{} is the intersection of \GgtwoF{} and \ODF{}, containing full \FOt{}.

We recall that the satisfiability problem for \ODF{} is undecidable \cite{HK14}. To regain decidability its \emph{uniform} restriction, \UF{},
was introduced in \cite{HK14}. Roughly speaking, a boolean combination of atoms is allowed in \UF{} if all of them use precisely the same set of variables;
the exceptions are atoms with one free variable and equalities, which may be used freely. 
See \cite{HK14} or \cite{KK14} for a formal definition and more details on \UF{}.

We will also be interested in the loosely guarded fragment, \LGF, the guarded negation fragment, \GNFO, and their one-dimensional and
tri-guarded variations. They will be introduced in Section \ref{s:var}.

\section{Finite model property} \label{s:fmp}

In this section we prove the finite model property for \GgtwoFonedim{} and obtain (essentially optimal) upper bounds on the size of  minimal models
of its satisfiable formulas, as well as of formulas of its interesting subfragments. 

We introduce a Scott-type normal form for \GgtwoFonedim{}. Given a \GgtwoFonedim{} formula $\phi$ we say that it is
in \emph{normal form} if it has the following shape
\begin{align} \label{f:bnf}
\bigwedge_{i \in I} \forall \bar{x} (\gamma_i(\bar{x}) \rightarrow \psi_i(\bar{x})) \wedge 
  \bigwedge_{i \in I'} \forall x \exists \bar{y} \psi'_i(x, \bar{y}) \wedge
	 \forall xy \psi''(x,y)
\end{align}
where $I, I'$ are some sets of indices, the
$\psi_i$, $\psi_i'$, and $\psi''$ represent arbitrary quantifier-free formulas, and for every $i$, $\gamma_i$ is a proper guard for $\psi_i$. We remark  that
we do not require guards in formulas of the form $\forall \bar{\exists}$, even if they contain more than two variables, as their presence there 
is inessential 
(cf.~Remark in \cite{Gra99}, p.~1725).
In a rather standard fashion one can show the following lemma.

\begin{lemma} \label{l:normalform}
There is a polynomial nondeterministic procedure, taking as its input a \GgtwoFonedim{} formula $\phi$ and
producing a normal form formula $\phi'$ (over an extended signature), such that
\begin{enumerate}[(i)]
\item if $\str{A} \models \phi$ for some structure $\str{A}$ then there is a run of the procedure producing a normal form  $\phi'$
such that $\str{A}' \models \phi'$ for some expansion $\str{A}'$ of $\str{A}$,
\item if the procedure has a run producing $\phi'$ and $\str{A}' \models \phi'$, for some $\str{A}'$, then $\str{A}' \models \phi$.
\end{enumerate}
Moreover, if $\phi$ is without equality then the procedure produces $\phi'$ without equality; if $\phi$ is in \GFonedim{} then
the last conjunct $\forall xy \psi''(x,y)$ is not present in $\phi'$.
\end{lemma}

Lemma \ref{l:normalform}  allows us, when dealing with decidability or complexity issues  and when considering 
the size of minimal models of formulas in \GgtwoFonedim{}, to restrict attention to normal form sentences. The part of this lemma
starting with `moreover'  will allow us to use it effectively for \GgtwoFonedim{} without equality and for \GFonedim.

Our normal form is similar to  normal form for \GF{} \cite{Gra99}. It  adapts the latter to the one-dimensional setting and extends it by the last type of conjuncts.
The conversion to normal form in \cite{Gra99} is deterministic, it however cannot be used directly in our case as it adds one free variable
to every subformula, which spoils one-dimensionality and may lead to unguarded subformulas with three variables.

Let $\phi$ be a normal form formula and $\str{A}$ its model. Take $a \in A$ and a conjunct $\psi=\forall x \exists \bar{y} \psi'_i(x, \bar{y})$ of $\phi$.
Let $\bar{b}$ be a tuple of elements of $\str{A}$ such that $\str{A} \models \psi'_i(a, \bar{b})$. Then $\str{A} \restr (\{ a \} \cup \bar{b})$ is called
a \emph{witness structure} for $a$ and $\psi$.

\begin{theorem} \label{t:fmp}
Every satisfiable formula $\phi$ in
\begin{enumerate}[(i)] 
\item \GgtwoFonedim{} (with equality) has a finite model of size bounded doubly exponentially in $|\phi|$.
\item \GgtwoFonedim{} without equality has a finite model of size bounded exponentially in $|\phi|$.
\item \GFonedim{} (with or without equality) has a finite model of size bounded exponentially in $|\phi|$.
\end{enumerate}
\end{theorem}

We concentrate on showing (i) and then obtain (ii) and (iii) as a corollary from the finite model construction presented.
Let $\phi$ be a normal form \GgtwoFonedim{} formula as in (\ref{f:bnf}), and denote $n=|\phi|$. Let us fix an arbitrary model $\str{A}$ of $\phi$. We
construct a bounded model $\str{B} \models \phi$. We mimic the scheme of the classical construction
from \cite{GKV97} showing an exponential model property for \FOt{}, in particular we adapt the notions of \emph{kings} and \emph{court}.
The details, however, are more complicated.

\smallskip\noindent
{\em Court.} We say that an element $a \in A$ is a \emph{king} if $\type{\str{A}}{a}$ is realized in $\str{A}$ only by $a$; $\type{\str{A}}{a}$
is then called \emph{royal}.
As in the case of \FOt{} kings are important as their duplication may be forbidden by formulas like $\forall xy (Px \wedge Py \rightarrow x=y)$.
Let $K \subseteq A$ be the set of kings of $\str{A}$. 
For each $a \in K$
and each $i \in I'$ choose a witness structure $\str{W}_{a,i}$ for $a$ and $\psi'_i$ in $\str{A}$. Let $C=K \cup \bigcup_{a,i} W_{a,i}$.
We call $\str{C}$ the \emph{court} of $\str{A}$. The court will be retained in $\str{B}$. Note that the number of elements in $C$ is bounded exponentially in $n$, and it that the structure $\str{C}$
can be described using exponentially many bits (the latter is true since the arity of all relation symbols is bounded by $n$). 
Note that $K$, and thus also $C$ may be empty.

\smallskip\noindent
{\em Pattern witness structures.}
For each non-royal element $a \in A \setminus K$ we say that the isomorphism type of the structure $\str{A} \restr (K \cup \{a \})$
is the  $\str{K}$-type of $a$. Note that from a $\str{K}$-type of an element one can infer its $1$-type, and that the number of the
$\str{K}$-types realized in $\str{A}$ is bounded doubly exponentially in $n$. Denote by $\AAA^\str{K}$ the set of $\str{K}$-types realized in $\str{A}$ by the elements of $A \setminus K$. Later, we will allow ourselves to use the notion of a $\str{K}$-type in a natural way also for other structures with a distinguished substructure $\str{K}$.
For each $\pi \in \AAA^{\str{K}}$ choose an element  $a$ having $\str{K}$-type $\pi$ in $\str{A}$ and for each $i \in I'$ choose
a witness structure $\str{W}_{\pi,i}$ for $a$ and $\psi'_i$. 
 Let $\str{W}^{*}_{\pi,i} = \str{W}_{\pi,i} \restr 
(W_{\pi,i} \setminus (K \cup \{ a \})$). 
For each $\pi \in \AAA^{\str{K}}$, $i \in I'$ and  $j=0,1,2$ let $\str{W}^*_{\pi, i, j}$ be a fresh  isomorphic
copy of  $\str{W}^*_{\pi, i}$. 

\smallskip\noindent
{\em Universe.}
We define the universe of  $\str{B}$ as follows $B:=C \cup \bigcup_{\pi, i, j} W^*_{\pi, i, j}$,
where $\pi$ ranges over $\AAA^{\str{K}}$, $i$ over $I'$ and  $j$ over $\{0,1,2\}$.
We emphasise that the sets $W^*_{\pi, i, j}$ are disjoint from $C$ and from each other. 
We retain in $\str{B}$ the structure on $C$ from $\str{A}$ 
and  for each $\pi, i, j$ we make   $\str{B} \restr (K \cup W^*_{\pi, i, j})$ isomorphic to $\str{A} \restr (K \cup W^*_{\pi, i})$. 
This, in particular, makes the $\str{K}$-type in $\str{B}$ of each element $b$ belonging to some $W^*_{\pi, i, j}$ identical with the 
$\str{K}$-type in $\str{A}$ of the counterpart of $b$ from the original substructure $\str{W}_{\pi,i}$.

\smallskip\noindent
{\em Witness structures for the court.}
Let us consider an element $c \in C \setminus K$, and denote by  $\pi$ its $\str{K}$-type in $\str{A}$. For every $i \in I'$ 
make  $\str{B} \restr (\{c \} \cup ({W}_{\pi,i} \cap K) \cup {W}^*_{\pi,i,0})$
isomorphic to $\str{W}_{\pi,i}$. 
This provides a witness structure for $c$ and $\psi'_i$ in $\str{B}$.
Note that a single such step (for fixed $c$ and $i$) consists in defining relations on tuples containing $c$, at least one
element of  ${W}^*_{\pi,i,0}$ and possibly some elements of $K$, since relations on other relevant tuples were defined in the desired way 
in step \emph{Universe}.
Note that no conflicts (attempts to set the same atom to both true and false) can arise, when we perform this step for some $c$ and $i$ and then for 
the same $c$ and some $i' \not=i$, because in the first case we define truth-values of relations only on tuples
containing some element from $W^*_{\pi,i,0}$, and in the second---only on tuples containing some element from $W^*_{\pi,i',0}$,
and $W^*_{\pi,i,0}$ is disjoint from $W^*_{\pi,i',0}$.
Finally, when we perform this step for some $c$, and then for some $c' \not=c$ no conflicts arise since in the first case we
define relations only on tuples containing $c$ but not $c'$ and in the second---only on tuples containing $c'$ but not $c$.

\smallskip\noindent
{\em Witness structures for the other elements.}
Consider now any element $b \in B \setminus C$. Assume it belongs to ${W}^*_{\pi', i', j'}$ and that $\pi$ is the $\str{K}$-type of
$b$ in $\str{B} \restr (K \cup \{ b \})$.
For each $i \in I'$ 
make the structure on $\{b  \} \cup (W_{\pi, i} \cap K) \cup W^*_{\pi, i, (j'+1 \mod{3})}$ isomorphic
to $\str{W}_{\pi, i}$. 
This provides a witness structure for $b$ and $\psi'_i$ in $\str{B}$.
Again, to do it we need to define relations on some tuples containing $b$ and some element of 
$W^*_{\pi, i, (j'+1 \mod{3})}$, and, due to our strategy, this  can be done without
conflicts. 

\smallskip\noindent
{\em Completing the structure.}
For any pair of distinct  elements $b,b' \in B$ whose $2$-type has not yet been defined in $\str{B}$
choose a pair of distinct elements $a, a'$ with $\type{\str{A}}{a}=\type{\str{B}}{b}$ and $\type{\str{A}}{a'}=\type{\str{B}}{b'}$, 
and set $\type{\str{B}}{b,b'}:=\type{\str{A}}{a,a'}$. An appropriate pair $a,a'$ exists even if $\type{\str{B}}{b}=\type{\str{B}}{b'}$ since at least one
of $b,b'$ has a non-royal type. 
For any tuple $\bar{b}$ of elements of $B$
containing at least three distinct elements, and any relation symbol $R$ of arity $|\bar{b}|$, if the truth-value
of $R(\bar{b})$ in $\str{B}$ has not yet been defined then set it to \emph{false}.

This finishes the definition of $\str{B}$.
Let us now estimate its size.
We can bound the number and the arity of relation symbols by $n=|\phi|$. Then the size of $K$
is bounded by the  number of possible $1$-types, $2^n$. 
The size of $C$ is bounded by $2^n \cdot n (n-1)$, as each element $a$ of $K$ may need at most $n$ witness structures each
of them containing (besides $a$) at most $n-1$ elements.
The number of possible  relations of arity at most $n$ on a a set of $2^n+1$ 
elements is bounded by $2^{{(2^n+1)}^n} \le 2^{2^{n^2+n}}$, thus the number of $\str{K}$-types is bounded by $(2^{2^{n^2+n}})^n
= 2^{n\cdot 2^{n^2+n}} \le 2 ^{2 ^ {n^2 + 2n}} \le 2 ^{2 ^ {2n^2}}$ (for $n>1$). Finally, we can bound the size of
$B$ by $2^n + 2^n \cdot n (n-1) + 3n (n-1) \cdot 2 ^{2 ^ {2n^2}}$, doubly exponentially in $n$.

Presently, we explain that
 $\str{B} \models \phi$. First note that for each $b \in B$ and each $i \in I'$ there is an appropriate witness structure:
if $b \in K$ then this witness structure is provided in $\str{C}$ which is a substructure of $\str{B}$. If $b \in C \setminus K$
or $b \in B \setminus C$ then a proper witness structure is provided explicitly either in step \emph{Witness structure for the court}
or, resp., \emph{Witness structures for the other elements}. Thus $\str{B}$ satisfies all conjuncts of $\phi$ of
the form $\forall x \exists \bar{y} \psi'_i(x, \bar{y})$.

Consider any conjunct $\forall \bar{x} (\gamma_i(\bar{x}) \rightarrow \psi_i(\bar{x}))$ of $\phi$ and a tuple of elements $\bar{b}$ such
that $\str{B} \models \gamma_i(\bar{b})$. If $\bar{b} \subseteq C$ or $\bar{b} \subseteq K \cup W^*_{\pi, i, j}$ for some $\pi, i, j$
then the structure on $\bar{b}$ was made an isomorphic copy of some substructure of $\str{A}$ in step \emph{Universe}. Otherwise $\bar{b}$ contains at least two distinct elements. In this case 
the structure on $\bar{b}$ was made an isomorphic copy of some substructure of $\str{A}$ either in one of the steps \emph{Witness structures for the court},
\emph{Witness structures for the other elements} or in step \emph{Completing the structure} (in this last subcase $\bar{b}$ contains precisely two distinct elements). Thus $\str{B} \models \psi_i(\bar{b})$. 
Finally, consider the conjunct $\forall xy \psi''(x,y)$ and take any pair $b, b' \in B$. Again, the structure on $\{ b, b' \}$ is
an isomorphic copy of a substructure of $\str{A}$ defined (at the latests) in step \emph{Completing the structure}.

\medskip

This finishes the proof of (i). To see (ii) and (iii) we first observe that in both cases every  satisfiable formula $\phi$
has a model without kings. 
Given a structure $\str{A}$ we define two new structures $2\str{A}$ and $2\str{A}^+$, each of them with universe 
 $A \times \{0,1\}$ and the substructures on $A \times \{0 \}$ and $A \times \{1 \}$ isomorphic to $\str{A}$.
In $2\str{A}$ we make these two copies of $\str{A}$ completely disjoint by setting the truth-value of $R(\bar{a})$ to \emph{false} 
for any $R$ and any tuple $\bar{a}$ (of the appropriate length) contained neither in $A \times \{0 \}$
nor $A \times \{1 \}$. 
In $2\str{A}^+$, for any tuple $\bar{a}$ contained neither in $A \times \{0 \}$
nor $A \times \{1 \}$ and for any relation symbol $R$ of arity $|\bar{a}|$, if this tuple contains at least three distinct elements then 
we also define $R(\bar{a})$ to be \emph{false}. If $\bar{a}$ contains just two distinct elements, say $(a,0)$ and $(a',1)$, then for any 
relation symbol $R$ or arity $|\bar{a}|$ set $R(\bar{a})$ \emph{true} iff   $\str{A} \models R(\bar{a} \restr 1)$ where
$\bar{a} \restr 1 $ is the projection of the elements of $\bar{a}$ on their first position.

Observations that if $\phi$ is without equality and $\str{A} \models \phi$ then 
$2\str{A}^+ \models \phi$, and that if $\phi$ is in \GFonedim{} (even with equality) and $\str{A} \models \phi$ then $2\str{A} \models \phi$
are routine. Of course our new models are without kings.
Starting our small model construction from a model without kings   we get $K=\emptyset$ and thus $\str{K}$-types trivialize
to $1$-types, which means that their number is bounded singly exponentially. Also $C = \emptyset$ and  thus we construct $\str{B}$ out of the
$W^*_{\pi, i, j}$ where $\pi$ ranges over the set of $1$-types, the number of possible $i$ is linear in $n$ and there
are just three possible values of $j$. The size of each $\str{W}^*_{\pi, i, j}$ is linear in $n$. The size of the constructed
models can be thus estimated by $3n(n-1) \cdot 2^n$. Hence part (ii) and (iii) of Thm.~\ref{t:fmp} hold.

\section{Complexity} \label{s:comp}

In this section we establish the complexity of the considered logics.

\begin{theorem} \label{t:complexities}
The satisfiability problem (= finite satisfiability problem)
\begin{enumerate}[(i)] 
\item for \GgtwoFonedim{} with equality is \TwoExpTime-complete.
\item for \GgtwoFonedim{} without equality is \NExpTime-complete.
\item for \GFonedim{} is \NExpTime-complete. 
\end{enumerate}
\end{theorem}

\smallskip\noindent
{\bf Upper bound in (i).}
We design an
alternating satisfiability test for \GgtwoFonedim{} using only exponential space. A \TwoExpTime-upper bound follows then from the fact
that \AExpSpace$=$\TwoExpTime{} (Chandra, Kozen, Stockmeyer \cite{CKS81}). The procedure takes as its input a \GgtwoFonedim{} formula $\phi$ 
and works as described below. 
For simplicity our description is slightly informal. In particular, we do not precisely specify how 
structures constructed during its execution are represented. We also allow ourselves to write ``guess an object $X$ such that $Y$''
instead of more accurate ``guess an object $X$; verify if $X$ meets property $Y$; if it does not then {\bf reject}''.

\begin{enumerate}
\item Nondeterministically compute a normal form $\phi'$ as in Lemma \ref{l:normalform}. Let $n:=|\phi'|$.
\item {\bf Guess} a set of $1$-types $\AAA = \AAA_{r} \;\dot{\cup} \;\AAA_{nr}$ over the signature of $\phi'$ (royal and non-royal types), 
such that for any  $\alpha_1$, $\alpha_2$ (possibly $\alpha_1=\alpha_2$)
such that $\alpha_1 \in \AAA$ and $\alpha_2 \in \AAA_{nr}$ there is a $2$-type $\beta$ such that  $\beta \restr x_1=\alpha_1$ and
$\beta \restr x_2=\alpha_2$, and $\beta$ does not violate the universal conjuncts of  $\phi'$.
\item {\bf Guess} structures $\str{K}$, $\str{C}$ of size at most $2^n$ and $2^n \cdot n^2$, resp., with $\str{K}$ being a substructure of $\str{C}$, such that 
(i) $\AAA[K]=\AAA_r$,
(ii) $\AAA[C\setminus K] \subseteq \AAA_{nr}$,
(iii) each element of $K$ has all the required witness structures for $\forall \bar{\exists}$ conjuncts of $\phi'$ in $\str{C}$, and
(iv) universal conjuncts of $\phi'$ are not violated in $\str{C}$.
\item {\bf Universally choose} an element $c \in C \setminus K$ and a conjunct $\psi$ of $\phi'$ of type $\forall \bar{\exists}$.
Set $\str{F}:=\str{C} \restr (K \cup \{c \})$.
\item Set $Counter:=0$.
\item 
{\bf Guess} an extension $\str{D}$ of $\str{F}$, with universe $D=K \cup \{c \} \cup \{  a_1, \ldots, a_t\}$, such 
that 
(i) $\type{\str{D}}{a_i} \in \AAA_{nr}$ for all $i$, 
 (ii) 
for some $k_1, \ldots, k_s \in K$ the structure $\str{W}=\str{D} \restr \{c, k_1, \ldots, k_s, a_1, \ldots, a_t\}$ is a
witness structure for $c$ and $\psi$,
(iii) universal conjuncts of $\phi'$ are not violated in $\str{D}$.
If $t=0$ then {\bf accept}.
\item {\bf Universally choose} a new value for $c$ from $\{a_1, \ldots, a_t \}$ and
a conjunct $\psi$ of $\phi'$ of the form $\forall \bar{\exists}$.  Set $\str{F}:=\str{F} \restr (K \cup \{c \})$.
\item $Counter:=Counter +1$
\item If $Counter < 2^{2^{2n^2}}$ then goto 6 else {\bf accept}.
\end{enumerate}

Let us first note that exponential space is sufficient to perform the above algorithm. By Lemma \ref{l:normalform} we have that $n$ is
bounded polynomially in $|\phi|$. The number of $1$-types in $\AAA$ is also bounded by $2^n$, as a $1$-type is determined by a
subset of the signature. For some pairs of $1$-types we need to guess a $2$-type whose description is exponential (there are at most 
$2^n$ tuples of length not greater than $n$ consisting of a pair of elements, and at most $n$ relation symbols). The size of the structure $\str{C}$ guessed in Step 4 is explicitly required to be exponential in $n$. Also its
description requires only exponentially many bits (recall that the arity of all relations is bounded by $n$). 
Analogously we can bound the size of structures 
$\str{D}$ guessed in Step 6. Finally, the value of $Counter$ is bounded doubly exponentially, so it also
can be written using exponentially many bits.

Now we argue that the procedure accepts its input $\phi$ iff $\phi$ is satisfiable.
Assume first that the procedure accepts $\phi$. We show that then $\phi'$ (and thus, by Lemma \ref{l:normalform}, also $\phi$) 
has a model. 
Consider an accepting run of the procedure. We may assume w.l.o.g. that this run is uniform, that is, when entering step 6, in configurations
differing only in the values of $Counter$ (but with isomorphic $\str{F}$s) it makes the same (isomorphic) guesses of $\str{D}$. Then the modification of this procedure in which Step 9 is replaced just by 'Goto 6'
can run infinitely (if necessary) without clashes. Indeed if the value $Counter=2^{2^{2n^2}}$ is reached we have a guarantee
that the $\str{K}$-type of the current $c$ appeared before in the computation (cf.~our estimations on the size of the small model constructed in the proof of Thm.~\ref{t:fmp}, in particular on the number of $\str{K}$-types).
We can construct a model for $\phi'$ starting from the substructure $\str{C}$ guessed in Step 4, and then providing witness structures for all conjuncts of the form $\forall\bar{\exists}$ of $\phi'$ and elements $c$ in accordance with guesses of $\str{D}$ is Step 6 (we add fresh copies of elements $a_1, \ldots, a_t$ and make the structure on the union of $K$, $\{c\}$ and the set of the newly added elements isomorphic to $\str{D}$). We complete 
the (usually infinite) structure as in Step \emph{Completing the structure} of the small model construction from the proof of Thm.~\ref{t:fmp} using
the $2$-types guaranteed in Step 1. As in that proof
we can also show that the constructed structure is a model of $\phi'$.

Conversely, assume that $\phi$ has a model $\str{A}^*$. Nondeterministically compute its normal form $\phi'$ and let $\str{A}\models \phi'$ 
be an expansion of $\str{A}^*$ guaranteed by Lemma \ref{l:normalform}. Let $\str{B}$ be a model of $\phi'$ constructed as in the proof of Thm.~\ref{t:fmp}, starting from $\str{A}$. W can now make all the guesses of our procedure in accordance with $\str{B}$: denoting $K_\str{B}$ and $C_\str{B}$ 
the set of kings and a court of $\str{B}$, resp.,
 we set $\AAA_r:=\AAA[K_\str{B}]$, $\AAA_{nr}:=\AAA[B \setminus K_\str{B}]$, 
$\str{K} := \str{K}_\str{B}$, $\str{C} := \str{C}_\str{B}$. Then in the loop 6-9, when a structure $\str{D}$ containing a witness
structure for $c$ and $\psi$ is going to be guessed we choose an element $c' \in \str{B}$ such that the $\str{K}$-types of $c'$ in $\str{B}$ and
$c$ in $\str{F}$ are identical and find a witness structure for $c'$ and $\psi$ in $\str{B}$. We set $\str{D}$ to be isomorphic
to the restriction of $\str{B}$ to the union of $K_\str{B}$ and this witness structure. This strategy naturally leads to acceptance.

\smallskip\noindent
{\bf Lower bound in (i).} 
We encode computations of an alternating Turing machine $M$ working in exponential space on its input $\bar{a}=a_{i_0}\ldots a_{i_{n-1}}$.

The general idea of the proof is not far from the ideas used in the proofs of the \TwoExpTime-lower bound for \GF{} \cite{Gra99} and \TwoNExpTime-lower bound for \GgtwoF{} with constants \cite{RS18}. We must, however, be careful to avoid quantification leaving more than one
variable free, which happens  in both the above-mentioned proofs.
\emph{E.g.}, in \cite{Gra99} configurations of a Turing machine are encoded by pairs of elements $a_1, a_2$; concretely, by the truth-values of some relations of arity $O(n)$ on tuples 
consisting of $a_1, a_2$. To enforce existence of successor configurations quantification leaving two free variables is needed there.

We assume that $M$ has states $s_0, s_1, \ldots, s_k$, where $s_0$ is the initial state, $s_{k-1}$ is the only accepting state, and
$s_k$ is the only rejecting state.  The alphabet of $M$ consists of letters $a_0, \ldots, a_l$ where $a_0$ represents \emph{blank}. Without loss of generality we assume that $M$ has precisely
two possible moves in every  configuration, that on its every computation path it enters the accepting or rejecting state no later than in $2^{2^n}$-th step, and then, after reaching such final state, does not stop but works infinitely in a trivial way, without changing its configuration.

For  $i=0, \ldots, k$ we use a predicate $S_i$, for  $i=0, \ldots, l$ we use a predicate $A_i$ and to describe the head position
we use a predicate $H$. Each of the $S_i$, $A_i$ and $H$ is of arity $1+n$.

We enforce the existence of two kings, called \emph{zero} and \emph{one}, marked, resp., by unary predicates $Z$ and $O$. They will also be called \emph{bits}, serve as
binary digits and will be used to encode the numbers of  tape cells.
\begin{gather} 
\exists x (Z(x) \wedge \neg O(x)) \wedge \forall xy (Z(x) \wedge Z(y) \rightarrow x=y) \label{eq:a}\\
\exists x (O(x) \wedge \neg Z(x)) \wedge \forall xy (O(x) \wedge O(y) \rightarrow x=y) \label{eq:b}
\end{gather}

The idea is that every element of a model encodes a configuration of $M$ in its relation to tuples of bits of size $n$.
Such a tuple of bits $\bar{b}$ can be naturally read as a number in the range  $[0, \ldots, 2^n-1]$. Let us think that $A_i(c,\bar{b})$ 
means that in the configuration encoded by $c$,  tape cell  $\bar{b}$ contains $a_i$, $H(c, \bar{b})$ denotes that
this tape cell is scanned by the head and, for a cell observed by the head, $S_i(c,\bar{b})$ means that $M$ is in state $s_i$.

To be able to speak about properties of configurations of $M$ in \GgtwoFonedim{} we introduce a predicate $C$ of arity $1+2n$, which will be made true at least for all
tuples consisting of an arbitrary element of a model followed by $2n$ bits. 
We first say that, for any $0 \le i < 2n$,  ${C}$ holds for some tuple consisting of $i$ ones and $2n-i$ zeros, and then propagate $C$
to all relevant tuples, using the fact that the pair of permutations $(2,1,3,\ldots, 2n)$ and $(2,3, \ldots, 2n,1)$ generates the whole permutation group $S_{2n}$. Below $\bar{z}=z_{2n-1}, z_{2n-2}, \ldots, z_1, z_0$.
\begin{align}   
\nonumber \forall x \exists t_1t_0  (O(t_1) \wedge  Z(t_0) \wedge & C(x,t_0, t_0, t_0, \ldots, t_0) \wedge \\
  \label{eq:c} &  C(x, t_1, t_0, t_0 \ldots, t_0) \wedge \\
\nonumber & C(x, t_1, t_1, t_0, \ldots, t_0) \wedge \ldots \wedge \\
 \nonumber &  C(x, t_1, t_1, t_1, \ldots, t_1) )
\end{align}
\vspace{-30pt}
\begin{align}
 \label{eq:d}\forall x \bar{z}  (C( & x,  \bar{z})   \rightarrow  C(x,z_{2n-2}, z_{2n-1}, z_{2n-3}, \ldots, z_0) \wedge  C(x, z_{2n-2}, z_{2n-3}, \ldots, z_0, z_{n-1}))
\end{align}
We use a convention that $\bar{u}, \bar{v}, \bar{w}$ are  tuples of variables of size $n$, 
$\bar{u}=u_{n-1}, \ldots, u_0$ and analogously for $\bar{v}$ and $\bar{w}$. 
We introduce abbreviations, $\lambdadiff(\bar{u}, \bar{v})$ and $\lambdanext(\bar{u}, \bar{v})$ for quantifier-free formulas
of size polynomial in $n$. The former is intended to say that the numbers encoded by $\bar{u}$ and $\bar{v}$ differ, the latter---that
the number encoded by $\bar{v}$ is greater by one than the number encoded by $\bar{u}$. \emph{E.g.}, $\lambdanext(\bar{u}, \bar{v})$ can
be defined as 
\begin{align}
  \bigvee\limits_{0 \le i < n} & (Z(u_i)  \wedge O(v_i) \wedge \bigwedge\limits_{j<i} (O(u_j) \wedge Z(v_j)) \wedge 
\bigwedge\limits_{j>i}(O(u_j) \leftrightarrow Z(v_j)))
\end{align}
Analogously, we use $\lambda^{i}(\bar{u})$ and $\lambda^{\ge i}(\bar{u})$ for formulas saying that the number encoded by $\bar{u}$
is, resp., equal to $i$ and  greater or equal $i$. Again, they can be defined in a standard way by quantifier-free, polynomially bounded formulas.

Now we ensure that every element properly encodes a configuration. The following formulas say that, resp.,  there is a tape cell scanned by the head, there is at most one such cell, this cell carries also information about the state, and
every tape cell contains precisely a single letter. Below $\dot{\bigvee}_i\psi_i$ is an easily definable shorthand for `exactly one of the $\psi_i$ holds'.
\begin{gather}
\label{eq:f}\forall x \exists \bar{u} (H(x,\bar{u}) \wedge \bigwedge_i (O(u_i) \vee Z(u_i))\\
\label{eq:g}\forall x\bar{u} \bar{v} ( C(x, \bar{u}, \bar{v}) \rightarrow H(x,\bar{u}) \wedge  \lambdadiff(\bar{u}, \bar{v}) \rightarrow \neg H(x,\bar{v}))\\
\label{eq:h}\forall x\bar{u} (H(x,\bar{u}) \rightarrow \dot{\bigvee}_i S_i(x,\bar{u}))\\
\label{eq:i}\forall x\bar{u} ( C(x, \bar{u}, \bar{u}) \rightarrow \dot{\bigvee}_i A_i(x,\bar{u}))
\end{gather}

We then say that every element has two successors, and, using the trick with permutations prepare appropriate guards. Predicates $Succ_i$ are of arity $2+3n$. For $i=1,2$ we write:
\begin{align}   
\nonumber \forall x \exists y t_1 t_0  (O(t_1) \wedge  Z(t_0) \wedge
  &  Succ_i(x, y,t_0, t_0, t_0, \ldots, t_0) \wedge \\
 \nonumber &  Succ_i(x,y, t_1, t_0, t_0 \ldots, t_0) \wedge \\
\label{eq:j}  & Succ_i(x, y, t_1, t_1, t_0, \ldots, t_0) \wedge \ldots  \wedge \\
 \nonumber &  Succ_i(x, y, t_1, t_1, t_1, \ldots, t_1) )
\end{align}
\vspace{-30pt}
\begin{align}
\nonumber \forall xy\bar{t} & (Succ_i( x,y,\bar{t})
  \rightarrow \\
		\label{eq:k}
& Succ_i(x,y,t_{3n-2}, t_{3n-1}, t_{3n-3}, \ldots, t_0)  \wedge Succ_i(x,y, t_{3n-2}, t_{3n-3}, \ldots, t_0, t_{3n-1}))
\end{align}

We next describe  the computations of $M$ on $\bar{a}$.
First we say that the letter at a tape cell not scanned by the head does not change in the successor configurations. 
For $i=1,2$:
\begin{align}
\label{eq:l}\forall xy\bar{u} ((Succ_i(x,y,&\bar{u}, \bar{u}, \bar{u})  \rightarrow \neg H(x,\bar{u}) \rightarrow \bigwedge_i( A_i(x,\bar{u}) \rightarrow A_i(y,\bar{u})))
\end{align}

Consider now existential moves. Assume that in an existential state $s_i$, reading a letter $a_j$ the machine has two possible
transitions: $(s_{i'}, a_{j'}, \rightarrow)$ 
and $(s_{i''}, a_{j''}, \leftarrow)$. Then we write:
\begin{align}
\nonumber  \forall & x  y\bar{u}\bar{v}\bar{w} (Succ_1(x,y,\bar{u}, \bar{v}, \bar{w}) \rightarrow   H(x,\bar{u}) \wedge S_i(x,\bar{u}) \wedge A_j(x,\bar{u})   \wedge \lambdanext(\bar{u}, \bar{v}) \wedge \lambdanext(\bar{w}, \bar{u}) \rightarrow \\
\label{eq:m}  & \;\;\;\;\;\;\;\;\; (H(y,\bar{v}) \wedge S_{i'}(y, \bar{v}) \wedge A_{j'}(y, \bar{u})) \vee (H(y,\bar{w}) \wedge S_{i''}(y, \bar{w}) \wedge
A_{j''}(y, \bar{u}))
                                       )
																			\end{align}

Similarly, assume that $M$ has moves as above in a universal state $s_i$. We write:
\begin{align}
\nonumber  \forall  xy & \bar{u} \bar{v} \bar{v}  (Succ_1(x,y, \bar{u}, \bar{v}, \bar{v}) \rightarrow \\
\label{eq:n} & H(x,\bar{u}) \wedge S_i(x,\bar{u}) \wedge A_j(x,\bar{u})   \wedge \lambdanext(\bar{u}, \bar{v}) \rightarrow  {H(y,\bar{v}) \wedge S_{i'}(y, \bar{v}) \wedge A_{j'}(y, \bar{u}) 
                                       )}
\end{align}
\vspace{-30pt}
\begin{align}
\nonumber \forall xy & \bar{u} \bar{w} \bar{w} (Succ_2(x,y,\bar{u}, \bar{w}, \bar{w}) \rightarrow\\
\label{eq:o} & H(x,\bar{u}) \wedge S_i(x,\bar{u}) \wedge A_j(x,\bar{u})    \wedge \lambdanext(\bar{w}, \bar{u}) \rightarrow  H(y,\bar{w}) \wedge S_{i'}(y, \bar{w}) \wedge A_{j'}(y, \bar{u}) 
                                       )
\end{align}

We finally say that a model does not contain
a configuration with the rejecting state and impose the existence of an element encoding the initial configuration.
\begin{gather}
\label{eq:p}\neg \exists x S_{k}(x) \wedge \exists x Init(x) 
\end{gather}
\vspace{-25pt}
\begin{align}
\nonumber \forall x \bar{u}  ( C  (x, \bar{u})   \rightarrow Init(x) \rightarrow  (&\lambda^{\sss =0}(\bar{u}) \rightarrow H(x,\bar{u}) \wedge S_0(x,\bar{u}) \wedge A_{i_0}(x,\bar{u})) \wedge \\
\label{eq:r} &(\lambda^{\sss =1}(\bar{u}) \rightarrow A_{i_1}(x, \bar{u})) \wedge \ldots \wedge \\ 
 \nonumber & (\lambda^{\sss =n-1}(\bar{u}) \rightarrow A_{i_{n-1}}(x, \bar{u})) \wedge \\
\nonumber  & (\lambda^{\sss \ge n}(\bar{u}) \rightarrow A_{0}(x, \bar{u}) )  ) 
\end{align}

Showing that $M$ accepts $\bar{a}$ iff the constructed formula has a model is routine.

\smallskip\noindent
{\bf Upper bounds in (ii) and (iii).}
In both cases we have proved an exponential model property. Thus, to test satisfiability it suffices to guess
an exponentially bounded structure and verify that it indeed is a model. 
More precisely, given
a formula $\phi$ we nondeterministically convert it into normal form $\phi'$.
We guess an exponentially bounded model $\str{B}$ of $\phi'$ (again we remark that not only the universe of
$\str{B}$ is bounded exponentially, but also the description
of $\str{B}$, since we are dealing only with at most $|\phi'|$ relations of arity at most $|\phi'|$),
and verify that it is indeed a model. The last task can be carried out in an exhaustive way: for each $b \in B$ and
each conjunct of $\phi'$ of the form $\forall x \exists \bar{y} \psi'_i(x,\bar{y})$ guess which elements form a witness structure for $b$ and this conjunct
and check that
they indeed form a required witness structure; 
for each conjunct $\forall \bar{x} (\gamma_i(\bar{x}) \rightarrow \psi_i(\bar{x}))$ 
enumerate all tuples $\bar{b}$ of elements of $B$ such that $|\bar{b}|=|\bar{x}|$  and check that $\str{B} \models \gamma_i(\bar{b})  \rightarrow \psi_i(\bar{b})$.
Proceed analogously with the conjunct $\forall xy \psi''(x,y)$.

\smallskip\noindent
{\bf Lower bounds in (ii) and (iii).} 
It suffices to show \NExpTime-lower bound for \GFonedim{} without equality. As advertised in the Introduction, we even strengthen this result using only uniform formulas, that is we show \NExpTime-hardness of the \emph{uniform} one-dimensional guarded fragment being the intersection of $\GF$ and $\UF$.  For our current purposes
it is sufficient to say that conjunctions of sentences $\exists \bar{x}  \psi(\bar{x})$ and $\forall \bar{x} \psi(\bar{x})$ with quantifier-free $\psi$ are uniform if all atoms of $\phi$ use either all variables of $\bar{x}$ or just one of them. We use only formulas of such kind. For a general definition of \UF{} see \cite{HK14} or \cite{KK14}. 
Our proof goes by an encoding of an exponential tiling problem and is given in the  Appendix.

\section{Variations on  extensions of the guarded fragment} \label{s:var}

Let us see what happens when the ideas of one-dimensionality, tri-guardedness and their combination are applied
to two extensions of the guarded fragment: the loosely guarded fragment, \LGF{}, introduced by van Benthem \cite{Ben97}, and the guarded negation fragment, \GNFO{},
introduced by B\'ar\'any, ten Cate and Segoufin 
\cite{BtCS15}.
\LGF{} is defined similarly to \GF{}, but the notion of the guard is more liberal: in subformulas of the form $\exists \bar{y} (\gamma(\bar{x}, \bar{y}) \wedge \phi(\bar{x}, \bar{y}))$ and
$\forall \bar{y} (\gamma(\bar{x}, \bar{y}) \rightarrow \phi(\bar{x}, \bar{y}))$ we do not require that $\gamma$ is atomic but allow it to be a conjunction of atoms such that for every variable from $\bar{y}$ and every variable from
$\bar{y} \cup \bar{x}$ there is an atom in $\gamma$ containing both of them. In \GNFO{} (atomic) guards are required not for
quantifiers but for negated subformulas. For a more detailed definition of \GNFO{} see \cite{BtCS15}.

\smallskip\noindent
{\bf One-dimensionality.}
First, let us see that the one-dimensionality decreases the complexity of \LGF{}, similarly as in  the case of \GF,  but does not affect the complexity of \GNFO{}.

\begin{theorem} \label{t:ext1}
\begin{enumerate}[(i)]
\item 
The satisfiability (= finite satisfiability) problem for the one-dimensional \LGF{}, \LGFonedim, is \NExpTime-complete. 
\LGFonedim{} has an exponential model property.
\item 
The satisfiability (= finite satisfiability) problem for the one-dimensional \GNFO{} is \TwoExpTime-complete.
\end{enumerate}
\end{theorem}

To prove (i) we adjust the small model construction from the proof of Thm.~\ref{t:fmp}, by using more copies of
witness structures and refining the strategy of providing witnesses.
The construction from the proof of Thm.~\ref{t:fmp} cannot be applied without any changes to the current scenario, as it may
accidentally form some cliques of cardinality greater than $2$ in the Gaifmann graph of the constructed model which then 
could work as loose guards and lead to a violation of some universal conjuncts of the input formula. 
Details are given in the Appendix.

To see (ii) note that \GNFO{} contains the unary negation fragment, \UNFO{}, whose satisfiability problem is already \TwoExpTime-hard.
\UNFO{} is not one-dimensional but can be polynomially translated to its equivalent UN-normal form (ten Cate, Segoufin \cite{StC13}), which is one-dimensional.
The upper bound is inherited from the upper bound for full \GNFO{} \cite{BtCS15}.

\smallskip\noindent
{\bf Tri-guardedness.}
Unfortunately, allowing for unguarded binary subformulas leads to undecidability already in the case of one-dimensional variants
of \LGF{} and \GNFO{}.

\begin{theorem}
The (finite) satisfiability problems for the one-dimensional \LGF{} or \GNFO{}, with unguarded subformulas with two variables, even without equality, are undecidable. 
\end{theorem}

In the case of \LGFonedim{}, unguarded binary subformulas give the power of full one-dimensional fragment \ODF.  Indeed by adding a conjunct
$\forall xy G^*(x,y)$ we would be able to guard any tuple of variables $x_1, \ldots, x_k$ by the conjunction $\bigwedge_{i \not=j} G^*(x_i, x_j)$. (A similar observation is present also in \cite{RS18}.) As the satisfiability problem for \ODF{} is undecidable \cite{HK14} this gives the undecidability of the considered variation of \LGF.
For the  one-dimensional \GNFO{}, using unguarded negations of binary atoms one can express transitivity of binary relations: $\neg \exists xyz (Rxy \wedge Ryz \wedge \neg Rxz)$. One-dimensional \GNFO{} contains the two-variable guarded fragment which becomes undecidable when extended by
transitive relations (Kiero\'nski \cite{Kie05}, Kazakov \cite{Kaz06}). Thus the claim follows.

\section{Adding constants} \label{s:constants}

Finally, we study the satisfiability problem for \GFonedim{} and \GgtwoFonedim{} with constants.
It turns out that in the presence of constants we  lose neither the decidability nor the finite model property, however, the complexity increases.
The following theorem completes Table \ref{tab:complexitiesconst}.

\begin{theorem} \label{t:constants}

\begin{enumerate}[(i)]
\item Every satisfiable formula in \GgtwoFonedim{} with constants has a finite model of size bounded doubly exponentially in its
length.
\item The satisfiability (= finite satisfiability) problem for \GFonedim{} with constants (with or without equality) is \TwoExpTime-complete.
\item The satisfiability (= finite satisfiability) problem for \GgtwoFonedim{} with constants (with or without equality) is \TwoNExpTime-complete.
\end{enumerate}
\end{theorem}

It is not difficult to see that Lemma \ref{l:normalform} holds for formulas with constants.
Thus, to show (i) we can use a minor adaptation of our small model construction from the proof of Thm.~\ref{t:fmp}. Indeed, 
interpretations of constants may be treated as kings. The number of $\str{K}$-types remains  doubly exponential. The construction
works then essentially without  changes, we only remark that in step \emph{Completing the structure}, when a $2$-type for a pair of
elements is chosen, we need to define the truth-values of all relations on tuples built out of these elements and
constants. This way we get a doubly exponential bound on the size of models.

The upper bound in (ii) follows from the fact that full \GF{} with constants is in \TwoExpTime{} \cite{Gra99}. 

The  upper bound in (iii) follows from the fact that full \GgtwoF{} with constants is
in \TwoNExpTime{} \cite{RS18}. We remark, however, that this  upper bound for \GgtwoF{} is obtained without proving the finite model property, thus 
to justify the upper bound for finite satisfiability of \GgtwoFonedim{} we must refer to part (i) of Thm.~\ref{t:constants}.

The corresponding lower bounds in (ii) and (iii) are proved in the Appendix.

\medskip\noindent
{\bf Acknowledgements.} The author would like to thank Sebastian Rudolph and the anonymous  reviewers  for  their helpful comments.

\bibliography{mybib}


\newpage
\appendix

\section{Normal form (Lemma \ref{l:normalform})} \label{a:normalform}
We sketch a proof of  Lemma \ref{l:normalform}.
\begin{proof}
Take a \GgtwoFonedim{} formula $\phi$. W.l.o.g.~assume that all its quantifiers are existential. We begin with an innermost subformula $\psi$ of $\phi$ starting with a block of quantifiers. 
If $\psi$ contains a free variable and a quantifier guard, \emph{i.e.}, it is of the form $\exists \bar{y} (\eta(x,\bar{y}) \wedge \psi'(x,\bar{y}))$, for some guard $\eta$, then we replace it by $P_\psi(x)$ and add two normal form
 conjuncts $\forall x   \exists \bar{y} (P_\psi(x) \rightarrow (\eta(x,\bar{y}) \wedge \psi'(x,\bar{y})) )$
and $\forall x\bar{y}  (\eta(x,\bar{y}) \rightarrow \neg \psi'(x,\bar{y}) \vee P_\psi(x))$ axiomatising $P_\psi$. 
If $\psi$ has a free variable but not a quantifier guard, \emph{i.e.}, it is of the form $\exists {y} \psi'(x,{y})$
we similarly replace it by $P_\psi(x)$ and add normal form conjuncts
 $\forall x \exists y (P_\psi(x) \rightarrow\psi'(x,y)) $
and $\forall xy   (\psi'(x,y) \rightarrow P_\psi(x))$. If $\psi$ is a subsentence,
\emph{i.e.}, it is of the form $\exists \bar{y} \psi'(\bar{y})$ then we 
nondeterministically guess its truth value, replace it
by $\top$ or $\bot$ according to this guess and add $\psi$ or, resp., $\neg \psi$ 
as a conjunct of our new formula.
Moving up the original formula $\phi$ we repeat this procedure for subformulas  that are now innermost, and so forth.
 The formula obtained in this process has, up to trivial logical transformations, the desired shape.
\end{proof}

\section{\NExpTime-lower bound for (uniform) GF${_1}$ (Thm.~\ref{t:complexities})} \label{a:lower1}
We proceed by a reduction from a variant of the tiling problem.
Let $\str{G}_{m}$ denote the standard grid on a finite $m \times m$ torus:
$\str{G}_m = ([0,m-1]^2,H,V )$,
$H = \set{((p, q), (p', q)): p' - p \equiv 1
\mod m}$, $V = \set{((p, q), (p, q')): q' - q \equiv 1 \mod m}$.  A
\emph{tiling system} is a quadruple $\cT = \langle C, c_0, Hor, Ver
\rangle$, where $C$ is a non-empty, finite set of \emph{colours},
$c_{0}$ is an element of $C$, and $Hor$, $Ver$ are binary relations on $C$
called the \emph{horizontal} and \emph{vertical} constraints,
respectively.  A \emph{tiling} for $\cT$ of a grid $\str{G}_m$ is a
function $f: [0,m-1]^2 \rightarrow C$ such that $f(0,0) = c_0$ and, for
all $d\in [0,m-1]^2$, the pair $\langle f(d), f(h(d)) \rangle$ is in $Hor$
and the pair $\langle f(d), f(v(d)) \rangle$ is in $Ver$.  The {\em exponential tiling problem} is defined as follows.
Given a number $n\in\N$ written in unary, and a tiling system $\cT$, verify
if $\cT$ has a tiling of the grid $\str{G}_m$, where $m={2^n}$.
It is well known that the exponential tiling problem is
$\NExpTime$-complete.

Given $n \in \N$ and a titling system $\cal T$ we now construct a formula satisfiable iff there is a tiling for $\cal T$ of the grid $\str{G}_m$ for $m=2^n$.
As in the proof of the lower bound in Thm.~\ref{t:complexities} (i), we mark two elements with predicates $Z$ and $O$. This time,
however, we cannot make them kings (formulas enforcing kings use both equality and unguarded subformulas with two variables). As we will see this will not be harmful.
Let $N$ be a predicate of arity $4n$. We say that two different elements exist, one in $Z$ and one in $O$.
We call them, resp., $zero$ and $one$. We enforce that $N$ holds for any tuple built out of zeros and ones. We use the trick
we already know:
we first say that, for any $0 \le i < 4n$,  $N$ holds for some tuple consisting of $i$ ones and $4n-i$ zeros, and then propagate $N$
to all relevant tuples.
Below $\bar{u}$ denotes the tuple of variables $u_1, \ldots, u_{4n}$.
\begin{align}   
\nonumber \exists xy & (Z(x) \wedge \neg O(x)  \wedge O(y) \wedge \neg Z(y) \wedge\\
  \nonumber  & \hspace{50pt} N(x,x,x,\ldots, x) \wedge \\
 \nonumber &  \hspace{50pt} N(y,x, x, \ldots, x) \wedge \\
\label{eq:start}  & \hspace{50pt} N(y,y,x, \ldots, x) \wedge\\
 \nonumber & \hspace{50pt}  \ldots  \wedge \\
 \nonumber &  \hspace{50pt} N(y,y,y, \ldots, y) 
\end{align}
\begin{align}
\label{eq:generb} \forall \bar{u} & (N(u_1,u_2, \ldots, u_{4n}) \rightarrow 
  N(u_2,u_1, u_3, \ldots, u_{4n}) \wedge N(u_2,u_3,\ldots, u_{4n}, u_1))
\end{align}

For every colour $t \in C$ we introduce a predicate $D_t$ of arity $4n$. 
A tuple of $zero$s and $one$s of length $n$ can be naturally interpreted as a number from the range $[0, \ldots , 2^n-1]$.
When $\bar{a}$, $\bar{b}$, $\bar{c}$, $\bar{d}$
are such tuples we want to interpret the fact that $D_t(\bar{a}, \bar{b}, \bar{c}, \bar{d})$ for some $\bar{c}$,
$\bar{d}$ holds as that the tile of colour $t$ is placed at coordinates $\bar{a}$, $\bar{b}$. 
To this end we  say that for every tuple
of length $4n$ precisely one of $D_t$ holds, and then we \emph{dummify} the second half of variables using the trick with permutations. Below
$\bar{w}$ represents the tuple $w_1, \ldots, w_{2n}$.
\begin{align}
\forall \bar{u} (N(\bar{u}) \rightarrow \overset{\cdot}{\bigvee_{t \in C}} D_t(\bar{u}))
\label{eq:plac}
\end{align}
\begin{align}
 \bigwedge_{t \in C} & (\forall \bar{x} \bar{y} \bar{w} (D_t(\bar{x}, \bar{y}, \bar{w}) \rightarrow    D_t(\bar{x}, \bar{y}, w_2, w_1, w_3, \ldots, w_{2n}) \wedge D_t(\bar{x}, \bar{y}, w_2,  \ldots, w_{2n}, w_1)    ))
\label{eq:cons}
\end{align}
Now we can easily encode horizontal and vertical constraints on tiles. Below we present an encoding of horizontal constraints. Vertical 
constraints can be encoded analogously.
\begin{align}
\label{eq:color}
 \bigwedge_{t \in C} & \big(\forall \bar{x} \bar{y} \bar{z} (D_t(\bar{x}, \bar{y}, \bar{z}, \bar{y}) \rightarrow 
\lambdanext(\bar{x}, \bar{z})  \rightarrow \bigvee_{t': (t,t') \in Hor} D_{t'}(\bar{z}, \bar{y}, \bar{x}, \bar{y})) \big)
 \end{align}
Recall that $\lambdanext(\bar{x}, \bar{z})$ is a quantifier-free, formula saying that 
$\bar{z}$ encodes the number greater by $1$ (we assume that $\lambdanext$ counts modulo $2^n$) than the one encoded by $\bar{x}$.

This finishes the reduction. Observing that the conjunction of (\ref{eq:start})--(\ref{eq:color}) (plus a formula for vertical constraints) is satisfiable iff $\cT$ tiles $\str{G}_m$ is, again, routine.

As remarked we used only uniform formulas, so we have the following corollary.

\begin{corollary}
The satisfiability problem for uniform one-dimensional guarded fragment is \NExpTime-complete.
\end{corollary}

\section{\NExpTime-upper bound for LGF$_1$ (Thm.~\ref{t:ext1})} \label{a:variations}

For \LGFonedim{} we can again use a Scott-like normal form, which now looks as follows:
\begin{align} \label{f:nf}
\bigwedge_{i \in I} \forall \bar{x} (\gamma_i(\bar{x}) \rightarrow \psi_i(\bar{x})) \wedge 
  \bigwedge_{i \in I'} \forall x \exists \bar{y} \psi'_i(x, \bar{y})
\end{align}
where the $\gamma_i(\bar{x})$ are \emph{loose guards}, in this case being  conjunctions of atoms such that every pair
of variables from $\bar{x}$ coincides in at least one atom. A natural counterpart of Lemma \ref{l:normalform}, with a similar proof,
 holds for \LGFonedim{}. Thus in the sequel we can restrict attention to normal form formulas of the shape as in (\ref{f:nf}).
 
To prove (i) we adjust the small model construction from the proof of Thm.~\ref{t:fmp}. We assume that $K=\emptyset$ (this can be
done w.l.o.g. since, as for \GF, if $\str{A} \models \phi$ then $2\str{A} \models \phi$, for any \LGF{} formula $\phi$). 
Thus, there are no kings,  $\str{K}$-types in $\str{A}$ become just $1$-types and there are exponentially many of them.

The construction from the proof of Thm.~\ref{t:fmp} cannot be applied without any changes to the current scenario, as it may
accidentally form some cliques of cardinality greater than $2$ in the Gaifmann graph of the constructed model which then 
could work as loose guards and lead to a violation of some universal conjuncts of (\ref{f:nf}). There are three sources 
from which such cliques may arise. 

First potential such source is the step \emph{Completing the structure} which in the proof of Thm.~\ref{t:fmp}
allows us to define the $2$-types not specified in the previous steps as any $2$-types from the original model $\str{A}$ which agree
with the $1$-types of the given elements.  In particular these $2$-types can contain some binary atoms.
The potential danger here can be easily avoided by removing from the $2$-types
which are going to be used in this step any non-unary atoms (which is a standard strategy in constructions of models for guarded formulas;
such a stragety  was not used in the proof of Thm.~\ref{t:fmp} since in that proof we needed to cover the case of \emph{tri}-guarded formulas).

Two other kinds of dangerous cliques could appear in our original construction for \GFonedim. Cliques of the first kind
are those created by three elements belonging to three different subsets $W^*_{\pi,i,j}$ with three different values of $j$. 
Consider, \emph{e.g.}, a normal form \LGFonedim{} formula $\forall xyz (Rxy \wedge Ryz \wedge Rzx \rightarrow \bot) \wedge \forall x \exists y Rxy$ and its model $\str{A}$ consisting of four elements joined by $R$ in a cyclic fashion. Starting from $\str{A}$, the construction from the proof of
Thm.~\ref{t:fmp} would construct a structure $\str{B}$ with three elements forming eventually a triangle forbidden by the universal conjunct.

Cliques of
the second kind could appear when two (or more) elements connected by some binary relation, say $R$, having the same type $\pi'$
and belonging to the same $W^*_{\pi,i,j}$
look for their witnesses for the $i'$-th conjunct $\forall\bar{\exists}$, since in such case the original strategy requires them to use the same $W^*_{\pi',i',j+1 \mod 3'}$, and, \emph{e.g.}, they both could connect by $R$ to the same element forming a triangle, which may
again be forbidden by some universal conjunct.

To avoid forming such problematic cliques we need to use more copies of each witness structure this time. For clarity let us describe the whole construction in 
details. Let $\phi$ be a normal form \LGFonedim{} formula as in (\ref{f:nf}) and $\str{A}$ its model without kings.

\smallskip\noindent
{\em Pattern witness structures.}
For each $1$-type (=$\str{K}$-type) $\pi$ realized in $\str{A}$ choose an element  $a \in A$ of $1$-type $\pi$ and for each $i \in I'$ choose
a witness structure $\str{W}_{\pi,i}$ for $a$ and $\psi'_i$. 
 Let $\str{W}^{*}_{\pi,i} = \str{W}_{\pi,i} \restr 
(W_{\pi,i} \setminus \{ a \}$). Let $n$ be the maximum size of $W^*_{\pi, i}$ across all the $\pi$ and $i$.

\smallskip\noindent
{\em Universe.}
We define the universe of  $\str{B}$ as follows $B:= \bigcup_{\pi, i, j, s} W^*_{\pi, i, j, s}$,
where $\pi$ ranges over all $1$-types realized in $\str{A}$,  $i$ over $I'$, $j$ over $\{0,1,2,3\}$, and
$s$ over $\{1, \ldots, n\}$. The sets $W^*_{\pi, i, j, s}$ are disjoint from each other. 
For all $\pi$, $i$, $j$, $s$ we make   $\str{B} \restr  W^*_{\pi, i, j, s}$ isomorphic to $\str{A} \restr W^*_{\pi, i}$.
Note that in comparison to the case of \GFonedim{} we have four possible values for the index $j$ instead of three and an additional index $s$.

\smallskip\noindent
{\em Providing witnesses.} 
Let us number the elements in each $W^*_{\pi, i,j,s}$ from $1$ up to, at most, $n$.
Consider now any element $b \in B$. Assume it belongs to ${W}^*_{\pi', i', j', s'}$, is numbered $s$, and has $1$-type
$\pi$. For each $i \in I'$ 
make the structure on $\{b  \} \cup  W^*_{\pi, i, (j'+1 \mod{4}),s}$ isomorphic
to $\str{W}_{\pi, i}$. 
This provides a witness structure for $b$ and $\psi'_i$ in $\str{B}$.
Note that there are no conflicts with the  previously defined substructure on 
$W^*_{\pi, i, (j'+1 \mod{4}),s}$.

\smallskip\noindent
{\em Completing the structure.}
For any tuple $\bar{b}$ of elements of $B$
containing at least two distinct elements, and any relation symbol $R$ of arity $|\bar{b}|$, if the truth-value
of $R(\bar{b})$ in $\str{B}$ has not yet been defined then set it to \emph{false}. Note that this step differs
from the corresponding step in the proof of Thm.~\ref{t:fmp}, as announced.

\smallskip 
This finishes the definition of $\str{B}$.
We remark that using four values for $j$ instead of three guarantees that dangerous cliques of the first kind will not appear. On the other hand, introducing 
the extra index $s$ for witness structures, and the described strategy of providing wintesses guarantee that cliques of the second kind
are avoided. 

Actually, it is readily verified that
the only cliques that can appear in the  Gaifmann graph of $\str{B}$ are those consisting of elements from the same witness structure for some element. As the structure on them
is copied from the original structure $\str{A}$ it follows that they cannot lead to a violation of the universal conjuncts of $\phi$.
As we explicitly take care of providing witness structures for all elements we get that $\str{B}$ is the desired exponential model of $\phi$.

\NExpTime-upper complexity bound then easily follows: we can just guess an exponentially bounded structure and verify that it is
indeed a model of $\phi$, similarly as described in the case of \GFonedim{} in the proof of Thm.~\ref{t:complexities} (iii).

\section{Complexity in the presence of constants (Thm.~\ref{t:constants})} \label{a:constants}

In this section we prove Thm.~\ref{t:constants}.

It is not difficult to see that Lemma \ref{l:normalform} holds for formulas with constants.

Thus, to show (i) we can use a minor adaptation of our small model construction from the proof of Thm.~\ref{t:fmp}. Indeed, 
interpretations of constants may be treated as kings. The number of $\str{K}$-types remains  doubly exponential. The construction
works then essentially without  changes, we only remark that in step \emph{Completing the structure}, when a $2$-type for a pair of
elements is chosen, we need to define the truth-values of all relations on tuples built out of these elements and
constants. This way we get a doubly exponential bound on the size of models.

The upper bound in (ii) follows from the fact that full \GF{} with constants is in \TwoExpTime{} \cite{Gra99}. 

For the upper bound in (iii) we design a simple algorithm which just converts a given formula into its normal form,
guesses its doubly exponentially bounded model guaranteed by part (i) of this lemma and verifies it. Alternatively,
the upper bound for the general satisfiability problem follows from the fact that full \GgtwoF{} with constants is
in \TwoNExpTime{} \cite{RS18}. We remark, however, that this  upper bound for \GgtwoF{} is obtained without proving the finite model property, thus it does not give automatically the upper bound
for finite satisfiability of \GgtwoFonedim{} with constants.

The rest of this section is devoted for lower bounds.

\subsection{\TwoExpTime-lower bound for GF$_1$ with constants}

Our \TwoExpTime-lower bound proof for \GgtwoFonedim{} with equality can be easily adapted to the case of \GFonedim{} with constants and
without equality. 

In the former, equality is needed only to enforce the existence of two kings (called \emph{bits}). 
Here their role will be played by two constants.

Moreover, a simple inspection
of the formulas we used shows that the formulas enforcing kings are the only formulas in which quantification in some subformulas with two variables is not guarded.

Thus, in the current scenario, we can use the proof for \GgtwoFonedim{} with only some minor changes: define bits using constants, replace existentially
quantified variables of (\ref{eq:c}) and (\ref{eq:j}) by constants, and remove 
formula (\ref{eq:f}). This formula is not necessary, since the existence of a tape cell scanned by the head in every configuration will be
enforced by requiring this explicitly in the initial configuration and then appropriately defining moves of the head as in formulas (\ref{eq:m})--(\ref{eq:o}).

For the reader's convenience we reproduce all the formulas
with all the required modifications below.

Formulas (\ref{eq:a})--(\ref{eq:b}) are replaced by:
\begin{gather}
Z(c_0) \wedge \neg O(c_0) \wedge O(c_1) \wedge \neg Z(c_1)
\end{gather}

Formula (\ref{eq:c}) is replaced by:
\begin{align}   
\nonumber \forall x  (
    &  C(x,c_0, c_0, c_0, \ldots, c_0) \wedge \\
 \nonumber &  C(x, c_1, c_0, c_0 \ldots, c_0) \wedge \\
  & C(x, c_1, c_1, c_0, \ldots, c_0) \wedge \ldots \wedge \\
 \nonumber &  C(x, c_1, c_1, c_1, \ldots, c_1) )
\end{align}

Formula (\ref{eq:d}) is retained:
\begin{align}
\nonumber & \forall x \bar{z} (C(x,\bar{z})\rightarrow\\ 
& \hspace{50pt} C(x,z_{2n-2}, z_{2n-1}, z_{2n-3}, \ldots, z_0) \wedge \\
\nonumber & \hspace{50pt}  C(x, z_{2n-2}, z_{2n-3}, \ldots, z_0, z_{n-1}))
\end{align}

Formula (\ref{eq:f}) is completely removed. Formulas (\ref{eq:g})--(\ref{eq:i}) are retained:
\begin{align}
&\forall x\bar{u} \bar{v} ( C(x, \bar{u}, \bar{v}) \rightarrow H(x,\bar{u}) \wedge  \lambdadiff(\bar{u}, \bar{v}) \rightarrow \neg H(x,\bar{v}))\\
&\forall x\bar{u} (H(x,\bar{u}) \rightarrow \dot{\bigvee}_i S_i(x,\bar{u}))\\
&\forall x\bar{u} ( C(x, \bar{u}, \bar{u}) \rightarrow \dot{\bigvee}_i A_i(x,\bar{u}))
\end{align}

Formula (\ref{eq:j}) is replaced by:
\begin{align}   
\nonumber \forall x \exists y  (&
   Succ_i(x, y,c_0, c_0, c_0, \ldots, c_0) \wedge \\
 \nonumber &  Succ_i(x,y, c_1, c_0, c_0 \ldots, c_0) \wedge \\
  & Succ_i(x, y, c_1, c_1, c_0, \ldots, c_0) \wedge\\
 \nonumber &  \ldots  \wedge \\
 \nonumber &  Succ_i(x, y, c_1, c_1, c_1, \ldots, c_1) )
\end{align}

Formula (\ref{eq:k}) is retained;
 \begin{align}
\nonumber \forall  xy\bar{t} & (Succ_i(x,y,\bar{t})  \rightarrow \\
& Succ_i(x,y,t_{3n-2}, t_{3n-1}, t_{3n-3}, \ldots, t_0)  \wedge \\
& \nonumber \hspace{10pt} Succ_i(x,y, t_{3n-2}, t_{3n-3}, \ldots, t_0, t_{3n-1}))
\end{align}

Formulas (\ref{eq:l})--(\ref{eq:r}) are retained:

\begin{align}
\label{eq:s}\forall xy\bar{u} ((Succ_i(x,y,&\bar{u}, \bar{u}, \bar{u})  \rightarrow\\
\nonumber  & \neg H(x,\bar{u}) \rightarrow \bigwedge_i( A_i(x,\bar{u}) \rightarrow A_i(y,\bar{u})))
\end{align}

\begin{align}
\nonumber  \forall & x  y\bar{u}\bar{v}\bar{w} (Succ_1(x,y,\bar{u}, \bar{v}, \bar{w}) \rightarrow  \\
\nonumber  & H(x,\bar{u}) \wedge S_i(x,\bar{u}) \wedge A_j(x,\bar{u})   \wedge \lambdanext(\bar{u}, \bar{v}) \wedge \lambdanext(\bar{w}, \bar{u}) \rightarrow \\
\label{eq:t}  & \;\;\;\;\; (H(y,\bar{v}) \wedge S_{i'}(y, \bar{v}) \wedge A_{j'}(y, \bar{u})) \vee \\
\nonumber & \hspace*{100pt}(H(y,\bar{w}) \wedge S_{i''}(y, \bar{w}) \wedge
A_{j''}(y, \bar{u}))
                                       )
																			\end{align}

\begin{align}
\nonumber  \forall  xy & \bar{u} \bar{v} \bar{v}  (Succ_1(x,y, \bar{u}, \bar{v}, \bar{v}) \rightarrow \\
\label{eq:u} & H(x,\bar{u}) \wedge S_i(x,\bar{u}) \wedge A_j(x,\bar{u})   \wedge \lambdanext(\bar{u}, \bar{v}) \rightarrow \\
\nonumber & \hspace{100pt} {H(y,\bar{v}) \wedge S_{i'}(y, \bar{v}) \wedge A_{j'}(y, \bar{u}) 
                                       )}
\end{align}
\begin{align}
\nonumber \forall xy & \bar{u} \bar{w} \bar{w} (Succ_2(x,y,\bar{u}, \bar{w}, \bar{w}) \rightarrow\\
\label{eq:w} & H(x,\bar{u}) \wedge S_i(x,\bar{u}) \wedge A_j(x,\bar{u})    \wedge \lambdanext(\bar{w}, \bar{u}) \rightarrow\\
 \nonumber & \hspace*{100pt}  H(y,\bar{w}) \wedge S_{i'}(y, \bar{w}) \wedge A_{j'}(y, \bar{u}) 
                                       )
\end{align}

\begin{gather}
\label{eq:p2}\neg \exists x S_{k}(x) \wedge \exists x Init(x) 
\end{gather}
\begin{align}
\nonumber \forall x \bar{u}  ( C & (x, \bar{u})   \rightarrow Init(x) \rightarrow \\
\nonumber & (\lambda^{\sss =0}(\bar{u}) \rightarrow H(x,\bar{u}) \wedge S_0(x,\bar{u}) \wedge A_{i_0}(x,\bar{u})) \wedge \\
\nonumber &(\lambda^{\sss =1}(\bar{u}) \rightarrow A_{i_1}(x, \bar{u})) \wedge\\
\label{eq:z2} &   \ldots \wedge \\ 
 \nonumber & (\lambda^{\sss =n-1}(\bar{u}) \rightarrow A_{i_{n-1}}(x, \bar{u})) \wedge \\
\nonumber  & (\lambda^{\sss \ge n}(\bar{u}) \rightarrow A_{0}(x, \bar{u}) )  ) 
\end{align}

\subsection{\TwoNExpTime-lower bound for TGF$_1$ with constants} \label{a:tgfonedim}

Our aim is now to define a doubly exponential toroidal grid, which can then serve to encode a doubly exponential tiling problem 
defined similarly as in Appendix \ref{a:lower1}.
This time, given a number $n\in\N$ written in unary, and a tiling system $\cT$, the problem is to verify
if $\cT$ has a tiling of the grid $\str{G}_m$, where $m=2^{2^n}$.
It is well known that the doubly exponential tiling problem is
$\TwoNExpTime$-complete. 

As in the previous proof we use two constants $c_0$ and $c_1$, whose interpretations will be, as usually, called \emph{bits}, or just \emph{zero} and \emph{one}. For unary predicates $O$ and $Z$ we say:
\begin{gather}
Z(c_0) \wedge \neg O(c_0) \wedge O(c_1) \wedge \neg Z(c_1)
\end{gather}

We introduce a predicate $G$ of arity $2+2n$, which will be made true for all tuples consisting of two arbitrary elements of a model followed by
$2n$ bits. Once more we use the trick similar to the one used to create the relation $C$ in the proof of Thm.~\ref{t:complexities} (i).
We
first say that, for any $0 \le i <2n$, $G$ holds for some tuple consisting of $i$ ones and $2n-i$ zeros and then propagate $G$
to all relevant tuples.
\begin{align}   
\nonumber \forall xy   (  & G(x,y,c_0, c_0, c_0, \ldots, c_0) \;\wedge\\
 \nonumber &  G(x,y, c_1, c_0, c_0 \ldots, c_0) \;\;\wedge \\
 \label{eq:z} & G(x,y, c_1, c_1, c_0, \ldots, c_0) \wedge  \\
 \nonumber &  \ldots  \wedge \\
 \nonumber &  G(x,y, c_1, c_1, c_1, \ldots, c_1) )
\end{align}
\vspace{-20pt}
\begin{align}
\nonumber \forall  x \bar{z} & ( G(x,z_{2n-1}, z_{2n-2}, z_{2n-3}, \ldots, z_0)\rightarrow\\  
    & G(x,y,z_{2n-2}, z_{2n-1}, z_{2n-3}, \ldots, z_0) \wedge G(x,y, z_{2n-2}, z_{2n-3}, \ldots, z_0, z_{n-1}))
\end{align}

The predicate $G$ can be used as a guard for any formula using two variables intended to be interpreted as two arbitrary elements and up to $2n$ variables intended
to be interpreted as bits. This turns out to be sufficient for our purposes. 

It is probably worth commenting that this kind of a `universal guard' cannot be enforced with in 
\GgtwoFonedim{} with equality (and thus with the ability of enforcing kings) but
without constants. The reason is that kings would need to be quantified in  (\ref{eq:z}) which would lead to a \emph{two}-dimensional
formula.

All the forthcoming formulas can be guarded 
by $G(x,y,\bar{u}, \bar{v})$, $G(x,x,\bar{u}, \bar{u})$, $G(x,x,\bar{u}, \bar{v})$ or $(x,y,\bar{u}, \bar{u})$, if necessary. For brevity, we omit
such guards in our exposition.

We endow each element with a pair of coordinates in the range $[0, 2^{2^n}-1]$, treating 
 tuples of length $n$ consisting of bits as indices of binary digits.
We use predicate $B_H$ of arity $n+1$ to encode the \emph{horizontal coordinate} of element $a$.
For a sequence of constants $\bar{c}$ of length $n$ we interpret the value of $B_H(a, \bar{c})$ as the value of the $c$-th bit of
the horizontal coordinate.  

Recall our $\lambdanext$ abbreviation. We use it to prepare ourselves for defining addition of $1$ to horizontal and
vertical coordinates (addition in the range $[0, \ldots, 2^{2^n}-1]$). To this end we divide positions of coordinates
into $Pivot$ (the least significant $0$), $Tail$ (positions to the right from the $Pivot$) and $Head$ (positions to the left from the $Pivot$).
\begin{gather}
\forall x \bar{u} (Tail_H(x,\bar{u}) \dot{\vee} Pivot_H(x, \bar{u}) \dot{\vee} Head_H(x, \bar{u}) )\\
\forall x (\neg B_H(x,c_0, \ldots, c_0) \rightarrow Pivot_H(x,c_0, \ldots, c_0))\\
\forall x (B_H(x,c_0, \ldots, c_0) \rightarrow Tail_H(x,c_0, \ldots, c_0))\\
\forall x \bar{u} \bar{v}  (\lambdanext(\bar{u}, \bar{v}) \wedge  B_H(x, \bar{v}) \wedge Tail_H(x,\bar{u}) \rightarrow Tail_H(x,\bar{v}))\\
\forall x \bar{u} \bar{v}  (\lambdanext(\bar{u}, \bar{v}) \wedge  \neg B_H(x, \bar{v}) \wedge Tail_H(x,\bar{u}) \rightarrow  Pivot_H(x,\bar{v}))\\
\forall x \bar{u} \bar{v}  (\lambdanext(\bar{u}, \bar{v}) \wedge (Pivot_H(x, \bar{u}) \vee Head_H(x, \bar{u})) \rightarrow  Head_H(x,\bar{v}) )
\end{gather}

We proceed analogously for vertical coordinates, using predicates $Tail_V$, $Pivot_V$, $Head_V$, and $B_V$.

We enforce the existence of the origo, and a horizontal and a vertical neighbour of each element.
\begin{gather}
\exists x ((\neg \exists \bar{u} B_H(x,\bar{u})) \wedge (\neg \exists \bar{u} B_V(x, \bar{u}))) \\
\forall x (\exists y Hxy \wedge \exists y Vxy)
\end{gather}

We now take care of the coordinates of the neighbouring elements. We say that the horizontal coordinates of elements connected by $H$
differs by one.
\begin{align}
\nonumber \forall xy \bar{u}  (Hxy \rightarrow  &(B_V(x,\bar{u}) \leftrightarrow B_V(y,\bar{u})) \wedge \\
  & (Tail_H(x,\bar{u}) \rightarrow \neg B_H(y,\bar{u})) \wedge \\
\nonumber	& (Pivot(x,\bar{u}) \rightarrow B_H(y, \bar{u}) \wedge \\
\nonumber	& (Head(x, \bar{u}) \rightarrow (B_H(y, \bar{u}) \leftrightarrow B_H(x, \bar{u}))) \\
\nonumber 	) &
\end{align}
We proceed analogously with vertical coordinates of elements connected by $V$.

We next connect any pair of elements with the same vertical and horizontal coordinates 
by binary predicate $Equal$. To this end we introduce an auxiliary predicate $EqUpTo_H$.
\begin{align}
\forall xy & ((B_H(x,c_0, \ldots, c_0) \leftrightarrow B_H(y,c_0, \ldots, c_0)) \rightarrow  EqualUpTo_H(x,y, c_0, \ldots, c_0))
\end{align}
\begin{align}
 \forall xy\bar{u}\bar{v} & ((\lambdanext(\bar{u}, \bar{v}) \wedge (B_H(x,\bar{v}) \leftrightarrow B_H(y,\bar{v})) \wedge  EqUpTo_H(x,y,\bar{u})) \rightarrow EqUpTo_H(x,y, \bar{v}))
\end{align}
Analogously we introduce $EqUpTo_V$, and write the advertised formula defining equality of the coordinates.
\begin{align}
\forall xy & ((EqUpTo_H(x,y,c_1, \ldots, c_1) \wedge EqUpTo_V(x,y,c_1, \ldots, c_1)) \rightarrow Equal(x,y))
\end{align}

Having such formulas we can now easily encode an instance of the doubly exponential tiling problem. In particular, the predicate $Equal$ may be used
to say that two elements having the same horizontal and vertical coordinates are tiled identically. We omit the routine details.

\end{document}